\documentclass[aps,pre,epsfigure,twocolumn]{revtex4-1}
\usepackage{graphicx}
\usepackage[colorlinks=true,linkcolor=blue,urlcolor=blue,citecolor=blue,pdfusetitle]{hyperref}
\usepackage[utf8]{inputenc}
\usepackage[english]{babel}
\usepackage{amsmath}
\usepackage[caption = false]{subfig}
\usepackage{graphicx,epstopdf}
\usepackage{graphicx}
\usepackage{amsmath}
\usepackage{blindtext}
\usepackage{lipsum}
\usepackage{amsmath}
\usepackage{amsfonts}
\usepackage{amssymb}
\usepackage{physics}
\usepackage{amsfonts}
\usepackage{pgfplots}
\pgfplotsset{compat=1.18}
\usepackage{bbm}
\usepackage{amssymb}
\usepackage{enumerate}
\usepackage{multirow}
\usepackage{color}
\usepackage{latexsym}
\usepackage{physics}
\usepackage{times,txfonts}

\newcommand{\etalpaper}[5]{#1, \emph{et al.}, #2 \textbf{#3}, #4 (#5).}
\newcommand{\bookentry}[3]{#1, \textit{#2} (#3).}

\begin{document}
\title{ Counter-diabatic control of open quantum battery charging}
\author{Sh. Ebrahimi$^{1}$}
\author{S. Salimi$^{1}$}
\email{shsalimi@uok.ac.ir}
\author{F. T. Tabesh$^{2}$}
\author{A. S. Khorashad$^{1}$}
\affiliation{$^{1}$Department of Physics, University of Kurdistan, P.O.Box 66177-15175, Sanandaj, Iran}
\affiliation{$^{2}$School of Quantum Physics and Matter, Institute for Research in Fundamental Sciences (IPM), P.O.Box 19395-5531, Tehran, Iran}
\begin{abstract}
Continuous-variable quantum batteries provide a viable platform for microscopic energy storage, yet environmental dissipation typically degrades both their charging efficiency and extractable work (ergotropy). In this work, we investigate the charging dynamics of an open continuous-variable quantum battery modeled as coupled harmonic oscillators driven by counter-diabatic (CD) protocols in the presence of a Markovian thermal reservoir. Using Williamson's theorem to track the Gaussian state evolution, we show that CD driving significantly enhances ergotropy compared to conventional unassisted driving. Specifically, in the underdamped regime, the CD protocol boosts the peak ergotropy while suppressing thermal degradation. In the overdamped regime, where conventional unassisted charging fails due to rapid thermal dissipation, the CD protocol maintains a robust, non-zero ergotropy. Furthermore, by evaluating the temporal fluence of the auxiliary control fields, we assess the energetic cost and confirm that these thermodynamic advantages are achieved with physically realistic control overheads. These results indicate that counter-diabatic driving provides a practical route to decoherence-resilient quantum energy storage.
\end{abstract}
\maketitle%
\frenchspacing
\section{Introduction}\label{sec:Introduction}
Quantum batteries are model systems used to investigate how energy can be stored and extracted by exploiting quantum phenomena, representing a fundamental departure from the constraints of classical electrochemical storage devices \cite{d1, d2}. Since their formal introduction by Alicki and Fannes \cite{d3}, quantum batteries have been shown to exploit resources such as entanglement, coherence, and many-body correlations; depending on the dissipation regime and driving protocol, these resources can enable faster charging times and enhanced work extraction compared to classical bounds \cite{FFS,d5,d6}. Theoretical investigations have demonstrated collective quantum advantages across various frameworks, such as Dicke batteries and coupled bosonic modes, where coherent operations boost charging power \cite{d8,FSM,d9,d10}. Among the proposed architectures, setups based on quantum harmonic oscillators have attracted considerable attention due to their analytical tractability and direct experimental viability \cite{d11}. In these models, often formulated as an interacting charger-battery pair (though reservoir-mediated architectures are also possible), the interplay between coupling strengths, driving fields, and frequency resonance largely determines the extractable work (ergotropy) \cite{d13,d14}.
Harmonic oscillator architectures have already seen proof-of-principle realizations across diverse platforms, notably superconducting LC circuits and trapped-ion systems, where fine control over bosonic modes facilitates fast charging protocols \cite{d15,d16}. These setups permit precise manipulation via external driving, making them testbeds for shortcuts to adiabaticity (STA) in dissipative regimes. STA encompass several quantum control methods: counter-diabatic (CD) driving, inverse engineering, and fast-forward scaling. These techniques suppress non-adiabatic transitions and enable quantum states to reach adiabatic targets within significantly shortened timescales \cite{d17,d18}. In quantum batteries, STA techniques offer a route to ultra-fast charging, which is especially critical in open-system settings where protracted adiabatic evolutions inevitably succumb to decoherence and environmental losses \cite{d19,d20}. Indeed, applying CD protocols to discrete and continuous architectures has been shown to protect and enhance extractable work \cite{d21}. Moreover, the experimental viability of STA has been validated across superconducting circuits \cite{d22}, nitrogen-vacancy (NV) centers \cite{d23,d24,d25}, and ultracold atomic ensembles \cite{d26,d27}.
While STA strategies are well-established for closed systems, extending them to open quantum dynamics poses fundamental conceptual and practical challenges. Current efforts largely follow two routes: adapting unitary protocols while mitigating dissipation through ancillary degrees of freedom or noise-resilient tailoring \cite{d30}, or directly formulating non-unitary shortcuts by embedding environmental coupling into the driving framework via non-Hermitian Hamiltonians \cite{d31,d32} or Lindbladian dynamics; the latter typically requires balanced gain and loss or tailored bath engineering. However, experimental implementation of open-system shortcuts often proves demanding, requiring tailored bath engineering or active reservoir feedback \cite{d34}.

To overcome these limitations, here we employ a unitary control strategy adapted to open-system dynamics to optimize the performance of an open quantum battery. The setup consists of a charger and a battery, both modeled as harmonic oscillators of frequency $\omega_0$, where the charger interacts with a Markovian thermal reservoir and is driven by an external coherent field, while the battery remains shielded from direct bath exposure. The open dynamics are solved rigorously via a Born--Markov master equation. We demonstrate that CD control yields significant performance enhancements over unassisted driving: in the underdamped regime, it suppresses thermal decoherence to provide an approximate seventy-fold amplification of peak ergotropy, while in the strongly overdamped regime, where standard charging fails due to rapid thermalization, it stabilizes the system and maintains non-vanishing extractable energy. A fluence-based energetic assessment confirms that these thermodynamic advantages are matched by the control costs, yielding up to two orders of magnitude improvement in resource efficiency.

The remainder of this paper is organized as follows. In Sec.~\ref{sec:model}, we introduce the coupled harmonic-oscillator battery setup and establish its Hamiltonian formulation. Section~\ref{sec:Ordinary_Driving} details the Born--Markov master equation governing the open dynamics under standard driving. In Sec.~\ref{sec:CD_dynamics}, we construct the CD driving protocol and analyze charging dynamics across both underdamped and overdamped regimes. Section~\ref{sec:energetic_cost} evaluates the ergotropy, control costs, and the resulting resource efficiency. Concluding remarks are provided in Sec.~\ref{sec:conclusion}.

\begin{figure}[tb]
    \centering
    \includegraphics[width=\columnwidth]{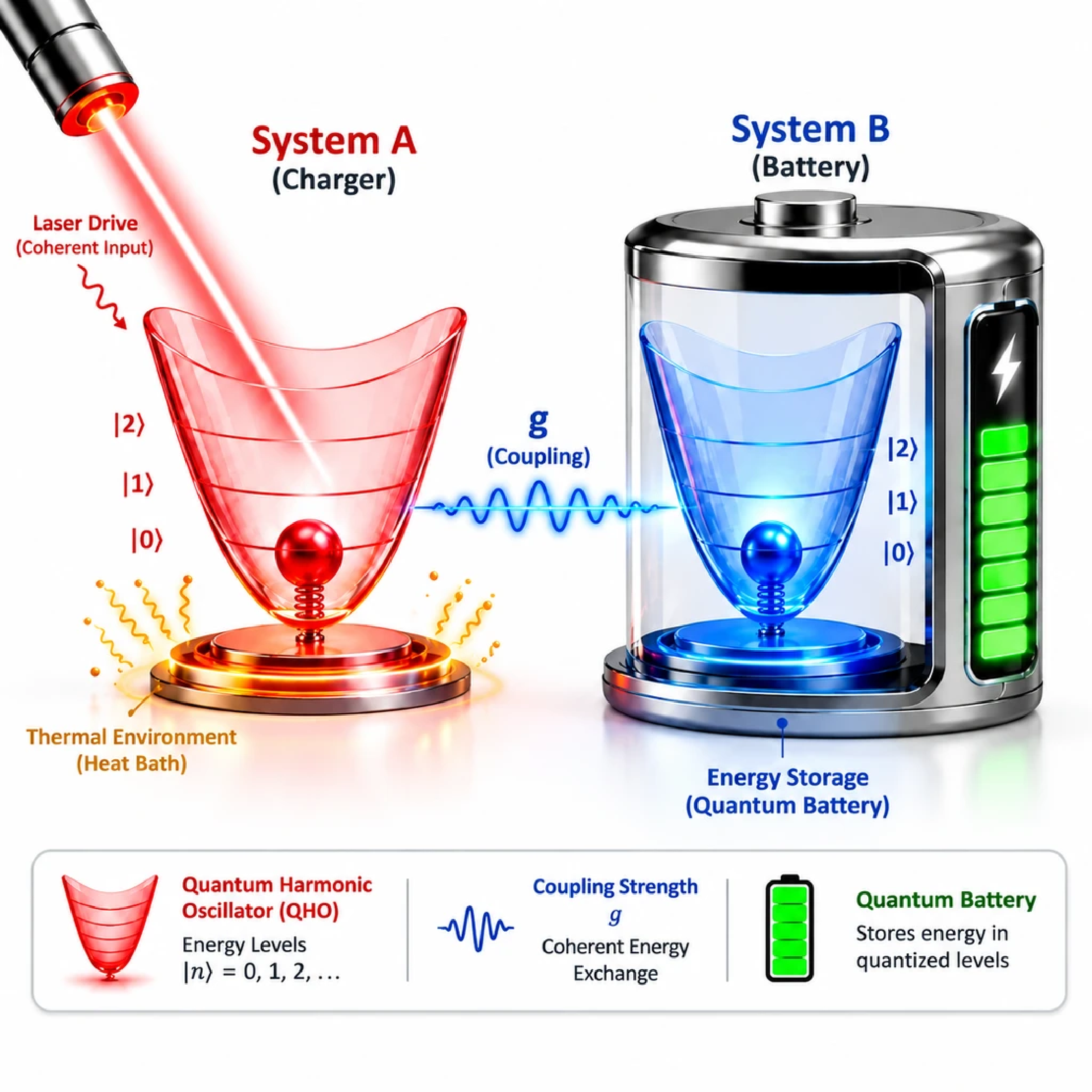}
    \caption{Schematic of the quantum battery (QB) charging protocol. Energy is supplied to charger $A$ through a coherent drive, while $A$ interacts with a thermal reservoir. The charging process is mediated by a time-dependent coupling $g(t)$ between $A$ and $B$, active over the interval $[0, \tau)$. Both the charger and the battery are modeled as quantum harmonic oscillators. A counter-diabatic driving field is applied to the charger to optimize energy transfer and stabilize the charging dynamics.}
    \label{fig:1}
\end{figure}

\section{Theoretical Framework and Physical Model}
\label{sec:model}

\subsection{System Hamiltonian and Control Protocol}

We consider a quantum charging architecture comprising two coupled quantum harmonic oscillators: a charger ($A$) and a battery ($B$), with bare frequencies $\omega_a$ and $\omega_b$, respectively. We work in natural units where $\hbar = 1$. A schematic of the setup, including the coherent driving of the charger, the time-dependent inter-oscillator coupling, and the thermal dissipation channel acting on the charger, is shown in Fig.~\ref{fig:1}.

The total Hamiltonian of the composite system in the laboratory frame is
\begin{equation}
    H_{\text{tot}}(t) = H_0 + H_{\text{drive}}(t) + H_{\text{int}}(t),
\end{equation}
where the free Hamiltonian is
\begin{equation}
    H_0 = \omega_a a^\dagger a + \omega_b b^\dagger b.
\end{equation}
Here, $a$ ($a^\dagger$) and $b$ ($b^\dagger$) denote the bosonic annihilation (creation) operators for the charger and battery modes, respectively, satisfying $[a, a^\dagger] = [b, b^\dagger] = 1$.

The charger is driven by an external field with carrier frequency $\omega_d$ and complex amplitude $\varepsilon(t)$. The driving Hamiltonian is
\begin{equation}
    H_{\text{drive}}(t) = -i \left[ \varepsilon(t) e^{-i\omega_d t} a^\dagger - \varepsilon^*(t) e^{i\omega_d t} a \right].
\end{equation}
Energy transfer between the charger and the battery is mediated by the bilinear coupling
\begin{equation}
    H_{\text{int}}(t) = g(t) \left( a^\dagger b + a b^\dagger \right) = g_0 \lambda(t) \left( a^\dagger b + a b^\dagger \right),
\end{equation}
where $g_0$ denotes the peak coupling strength and $\lambda(t)$ is the switching function:
\begin{equation}
    \lambda(t) =
    \begin{cases}
        1, & 0 \le t < \tau, \\
        0, & \text{otherwise}.
    \end{cases}
\end{equation}

To eliminate explicit fast oscillations at frequency $\omega_d$, we transform to the rotating frame via the unitary operator
\begin{equation}
    U(t) = \exp \left[ -i \omega_d t (a^\dagger a + b^\dagger b) \right].
\end{equation}
Within the standard rotating-wave approximation (RWA), the effective Hamiltonian reads
\begin{equation}
    H_{\text{eff}}(t) = \Delta_a a^\dagger a + \Delta_b b^\dagger b - i \left[ \varepsilon(t) a^\dagger - \varepsilon^*(t) a \right] + g_0 \lambda(t) \left( a^\dagger b + a b^\dagger \right),
\end{equation}
where $\Delta_a = \omega_a - \omega_d$ and $\Delta_b = \omega_b - \omega_d$ are the detunings. We define the relative detuning as $\Delta' \equiv \Delta_a - \Delta_b = \omega_a - \omega_b$. In the fully resonant case, $\omega_a = \omega_b = \omega_d \equiv \omega_0$, both detunings vanish.

\subsection{Open-System Dynamics}
\label{subsec:open_dynamics}

In a realistic setting, the charger is coupled to a thermal environment at temperature $T$, whereas the battery remains effectively isolated during charging. Under the Born--Markov and weak-coupling approximations, the open dynamics of the bipartite system is described by a Gorini--Kossakowski--Sudarshan--Lindblad (GKSL) master equation in the interaction picture \cite{d35,d36,d37}. We define the interaction-picture density operator with respect to
\begin{equation}
    H_0^{(\mathrm{rot})} = \Delta_a a^\dagger a + \Delta_b b^\dagger b
\end{equation}
as
\begin{equation}
    \tilde{\rho}_{AB}(t) = e^{i H_0^{(\mathrm{rot})} t} \rho_{AB}(t) e^{-i H_0^{(\mathrm{rot})} t}.
\end{equation}
In this picture, the transformed driving and coupling Hamiltonians become
\begin{equation}
    \tilde{H}_{\mathrm{drive}}(t) = -i \left[ \varepsilon(t) e^{i\Delta_a t} a^\dagger - \varepsilon^*(t) e^{-i\Delta_a t} a \right],
\end{equation}
and
\begin{equation}
    \tilde{H}_{\mathrm{int}}(t) = g_0 \lambda(t) \left[ a^\dagger b\, e^{i\Delta' t} + a b^\dagger e^{-i\Delta' t} \right],
\end{equation}
respectively. The reduced density operator $\tilde{\rho}_{AB}(t)$ then evolves according to
\begin{equation}
\begin{aligned}
    \frac{d\tilde{\rho}_{AB}(t)}{dt} ={}& -i \Big[ \tilde{H}_{\mathrm{drive}}(t) + \tilde{H}_{\mathrm{int}}(t),\, \tilde{\rho}_{AB}(t) \Big] \\
    & + \mathcal{L}_{\mathrm{loss}}\big(\tilde{\rho}_{AB}(t)\big) + \mathcal{L}_{\mathrm{gain}}\big(\tilde{\rho}_{AB}(t)\big).
\end{aligned}
\label{eq:master-equation}
\end{equation}
The dissipative superoperators describing photon loss and thermal excitation of the charger are
\begin{equation}
    \mathcal{L}_{\mathrm{loss}}(\rho) = \gamma(n_{\mathrm{th}} + 1) \left( a \rho a^\dagger - \frac{1}{2} \{a^\dagger a, \rho\} \right),
\end{equation}
and
\begin{equation}
    \mathcal{L}_{\mathrm{gain}}(\rho) = \gamma n_{\mathrm{th}} \left( a^\dagger \rho a - \frac{1}{2} \{a a^\dagger, \rho\} \right),
\end{equation}
where $\gamma$ is the decay rate of the charger and
\begin{equation}
    n_{\mathrm{th}} = \frac{1}{e^{\omega_0 / (k_B T)} - 1}
\end{equation}
is the mean thermal occupation of the reservoir evaluated at reference frequency $\omega_0$.

Initially, both charger and battery are prepared in their vacuum states,
\begin{equation}
    \rho_{AB}(0) = |0\rangle_A \langle 0|_A \otimes |0\rangle_B \langle 0|_B.
\end{equation}
At $t = 0$, the coherent drive injects energy into the charger, while the inter-oscillator interaction enables coherent energy transfer to the battery over $0 \le t < \tau$. During this process, dissipation and thermal fluctuations act exclusively on the charger mode.

Because master equation~\eqref{eq:master-equation} is at most quadratic in the bosonic operators and the initial state is Gaussian, Gaussianity is preserved throughout the evolution. Consequently, the reduced state of the battery mode, $\rho_B(\tau) = \operatorname{tr}_A [\tilde{\rho}_{AB}(\tau)]$, remains a single-mode Gaussian state for all $\tau$. The dynamics is therefore entirely determined by a closed set of coupled differential equations for the first- and second-order statistical moments, without requiring direct integration of the full density operator (see Appendix~\ref{app:equations_of_motion} for the explicit moment equations).

Under these conditions, we assess the charging performance and work extraction via two thermodynamic figures of merit: the total stored energy and the ergotropy \cite{d38}. The mean energy stored in the battery mode ($H_B = \omega_b b^\dagger b$) at the end of the charging interval $t = \tau$ is
\begin{equation}
    E_B(\tau) = \operatorname{tr}\!\left[ H_B \rho_B(\tau) \right] = \omega_b \langle b^\dagger b \rangle_\tau.
    \label{eq:stored_energy}
\end{equation}
The ergotropy $\mathcal{E}_B(\tau)$, defined as the maximum work extractable from $\rho_B(\tau)$ via cyclic local unitaries $U_B$ acting on mode $B$ \cite{d38,d40}, is the difference between the state energy and that of its passive counterpart $\rho_B^{(p)}(\tau)$:
\begin{equation}
\begin{aligned}
    \mathcal{E}_B(\tau) &= E_B(\tau) - \min_{U_B} \operatorname{tr}\!\left[ H_B U_B \rho_B(\tau) U_B^\dagger \right] \\
    &= E_B(\tau) - E_B^{(p)}(\tau) = \omega_b \left( \langle b^\dagger b \rangle_\tau - n_B^{(p)}(\tau) \right).
\end{aligned}
\label{eq:ergotropy_def}
\end{equation}
For single-mode Gaussian states, the passive state is uniquely determined by the symplectic invariant of the covariance matrix $\boldsymbol{\sigma}_B$. The extractable work is thus directly evaluated from the moments as
\begin{equation}
\mathcal{E}_B(\tau) = \omega_b \left[ \langle b^\dagger b \rangle - \frac{\sqrt{\mathcal{M}} - 1}{2} \right]_{t=\tau},
\label{eq:ergotropy_battery}
\end{equation}
where $\mathcal{M} \equiv 4 \det \boldsymbol{\sigma}_B$ takes the explicit form
\begin{equation}
\mathcal{M} = \left( 1 + 2\langle b^\dagger b \rangle - 2|\langle b \rangle|^2 \right)^2 - 4 \left| \langle b^2 \rangle - \langle b \rangle^2 \right|^2.
\label{eq:det_M}
\end{equation}
Here, $n_B^{(p)}(\tau) = (\sqrt{\mathcal{M}} - 1)/2$ quantifies the non-extractable, incoherent excitations accumulated in the battery mode (see Appendix~\ref{app:ergotropy_proof} for the phase-space derivation).
\subsection{Energy Contribution and Ergotropy Analysis}
\label{sec:ergotropy}

Because both charger $A$ and battery $B$ in our model are harmonic oscillators governed by linear equations of motion, the contributions stemming from the thermal reservoir and the coherent driving field decouple identically. For any system observable $\hat{\mathcal{O}}$, we introduce the expectation response functional
\begin{equation}
    \mathcal{R}_{\hat{\mathcal{O}}}(t; F, T) \equiv \operatorname{tr}\!\left[ \hat{\mathcal{O}} \tilde{\rho}_{AB}(t; F, T) \right].
    \label{eq:response_def}
\end{equation}
As proven in Appendix~\ref{app:proof_decomposition} via the linearity of the underlying moment dynamics, this response decomposes into thermal and coherent contributions:
\begin{equation}
    \mathcal{R}_{\hat{\mathcal{O}}}(t; F, T) = \mathcal{R}_{\hat{\mathcal{O}}}^{\mathrm{th}}(t; T) + \mathcal{R}_{\hat{\mathcal{O}}}^{\mathrm{coh}}(t; F),
    \label{eq:decomp_general}
\end{equation}
where $\mathcal{R}_{\hat{\mathcal{O}}}^{\mathrm{th}}(t; T) \equiv \operatorname{tr}[ \hat{\mathcal{O}} \tilde{\rho}_{AB}(t; 0, T) ]$ and $\mathcal{R}_{\hat{\mathcal{O}}}^{\mathrm{coh}}(t; F) \equiv \operatorname{tr}[ \hat{\mathcal{O}} \tilde{\rho}_{AB}(t; F, 0) ]$.

For the battery Hamiltonian $H_B = \omega_b b^\dagger b$, the total accumulated energy at the final charging time $\tau$, defined as $E_B(\tau; F, T) \equiv \mathcal{R}_{H_B}(\tau; F, T)$, inherits this linear splitting from Eq.~\eqref{eq:decomp_general}:
\begin{equation}
    E_B(\tau; F, T) = E_B^{\mathrm{th}}(\tau; T) + E_B^{\mathrm{coh}}(\tau; F),
    \label{eq:energy_decomp}
\end{equation}
with $E_B^{\mathrm{th}}(\tau; T) = \omega_b \mathcal{R}_{b^\dagger b}^{\mathrm{th}}(\tau; T)$ and $E_B^{\mathrm{coh}}(\tau; F) = \omega_b \mathcal{R}_{b^\dagger b}^{\mathrm{coh}}(\tau; F)$.

The ergotropy of the quantum battery represents the maximum work extractable via cyclic unitary operations and is given by
\begin{equation}
    \mathcal{E}_B(\tau; F, T) \equiv E_B(\tau; F, T) - E_B^{(p)}(\tau; F, T),
    \label{eq:ergotropy_def_split}
\end{equation}
where $E_B^{(p)}(\tau; F, T)$ denotes the energy of the passive state associated with the reduced battery state at time $\tau$. For a single-mode Gaussian state, the passive energy is determined solely by the determinant of the covariance matrix, $\mathcal{M}$, which characterizes the quantum and thermal fluctuations around the mean field~\cite{d39}. Since the coherent driving acts purely as a phase-space displacement operator, it shifts the first moments while leaving the covariance matrix and its invariant $\mathcal{M}$ unaffected. Consequently, $\mathcal{M}$ is independent of the coherent drive amplitude $F$ and is dictated by the thermal population:
\begin{equation}
    \mathcal{M} = \left[ 1 + 2 \mathcal{R}_{b^\dagger b}^{\mathrm{th}}(\tau; T) \right]^2.
    \label{eq:symplectic_det}
\end{equation}
Substituting Eq.~\eqref{eq:symplectic_det} into the expression for the passive energy yields
\begin{equation}
\begin{aligned}
    E_B^{(p)}(\tau; F, T) &= \omega_b \frac{\sqrt{\mathcal{M}}-1}{2} \\
    &= \omega_b \mathcal{R}_{b^\dagger b}^{\mathrm{th}}(\tau; T) = E_B^{\mathrm{th}}(\tau; T).
\end{aligned}
\label{eq:passive_energy}
\end{equation}

Equation~\eqref{eq:passive_energy} reveals that the energy associated with the thermal population is purely passive and useless for unitary work extraction. Substituting Eq.~\eqref{eq:passive_energy} and Eq.~\eqref{eq:energy_decomp} into Eq.~\eqref{eq:ergotropy_def_split}, the thermal contribution cancels out completely, leaving the ergotropy uniquely dictated by the coherent drive:
\begin{equation}
    \mathcal{E}_B(\tau; F, T) = E_B^{\mathrm{coh}}(\tau; F) = \omega_b \mathcal{R}_{b^\dagger b}^{\mathrm{coh}}(\tau; F), \quad \forall\, T \geq 0.
    \label{eq:ergotropy_final}
\end{equation}
In particular, when the system is driven solely by the thermal bath ($F=0$), the coherent contribution vanishes, leading to identically zero ergotropy ($\mathcal{E}_B(\tau; 0, T) = 0$).

These results establish a key physical insight: although charger $A$ is coupled to both a thermal bath and an external coherent drive, the extractable work stored in the battery originates exclusively from the coherent field, whereas the thermal bath contributes solely passive energy. Therefore, in the subsequent charging protocol analyses, it is sufficient to evaluate the coherent contribution to the battery energy, $\mathcal{E}_B(\tau) \equiv E_B^{\mathrm{coh}}(\tau; F)$.

\section{Ordinary Driving Protocol}
\label{sec:Ordinary_Driving}

We first analyze the charging dynamics of the quantum battery using a standard driving protocol, where the charger is subject to a smooth time-dependent coherent pulse envelope:
\begin{equation}
\varepsilon(t) = \varepsilon_0 \sin^2(\omega_d t), \quad \omega_d = \frac{\pi}{2\tau},
\label{eq:pulse_envelope}
\end{equation}
where $\tau$ represents the charging time interval $t \in [0, \tau]$. Compared to conventional square pulses, this pulse profile mitigates non-adiabatic transitions and reduces energetic expenditure. The evaluation and optimization of the driving cost are detailed in Appendix~\ref{app:driving-cost}. To quantify the energetic overhead of the external control over $[0, \tau]$, we employ the standard fluence functional~\cite{Werschnik2007}:
\begin{equation}
    C_d = \int_0^\tau \varepsilon^2(t)\, dt,
    \label{eq:driving_cost}
\end{equation}
capturing the time-integrated intensity of the external field.

To evaluate the system dynamics, we transform the first-order expectation values into the rotating frame by defining the transformed amplitudes $A(t) = \langle a \rangle e^{-i \Delta_a t}$ and $B(t) = \langle b \rangle e^{-i \Delta_b t}$. Using the coupled equations of motion for the first-order moments alongside the vacuum conditions $A(0) = B(0) = 0$, the auxiliary battery amplitude $B(t)$ can be integrated and substituted into the charger's equation of motion. Consequently, the coherent amplitude of the charger satisfies the driven damped oscillator equation:
\begin{equation}
    \ddot{A}(t) + \alpha \dot{A}(t) + \beta A(t) = -\dot{\varepsilon}(t) - i\Delta_b \varepsilon(t),
    \label{eq:charger_amplitude}
\end{equation}
where the complex coefficients $\alpha$ and $\beta$ encode the dissipation and detuning parameters:
\begin{equation}
    \alpha = \frac{\gamma}{2} + i(\Delta_a + \Delta_b), \qquad \beta = g^2 + i\frac{\gamma}{2}\Delta_b - \Delta_a \Delta_b.
    \label{eq:alpha_beta_def}
\end{equation}

Given the initial conditions $A(0) = \dot{A}(0) = 0$, the exact analytical solution for $A(t)$ is
\begin{equation}
    A(t) = e^{-\frac{\alpha}{2}t} \left( C_+ e^{\Omega t} + C_- e^{-\Omega t} \right) + C_0 + C_1 \cos(2\omega_d t) + C_2 \sin(2\omega_d t),
    \label{eq:A_t_solution}
\end{equation}
where $\Omega = \sqrt{\alpha^2/4 - \beta}$ and $D = -4\omega_d^2 + \beta$. The particular solution coefficients are given by
\begin{equation}
\begin{aligned}
    C_0 &= -\frac{i\Delta_b \varepsilon_0}{2\beta}, \\
    C_1 &= \frac{2\omega_d^2 \varepsilon_0 \alpha + \frac{i}{2} D \Delta_b \varepsilon_0}{D^2 + 4\omega_d^2 \alpha^2}, \\
    C_2 &= \frac{-\omega_d \varepsilon_0 D + i \alpha \omega_d \Delta_b \varepsilon_0}{D^2 + 4\omega_d^2 \alpha^2},
\end{aligned}
\label{eq:C_coefficients}
\end{equation}
and the integration constants $C_\pm$ are determined by the boundary conditions:
\begin{equation}
    C_+ = -(C_0 + C_1 + C_-), \qquad C_- = \frac{-(C_0 + C_1)\left(-\frac{\alpha}{2} + \Omega\right) + 2\omega_d C_2}{2\Omega}.
\end{equation}

With $A(t)$ determined, the coherent amplitude of the battery mode follows from direct integration:
\begin{equation}
    \langle b \rangle_\tau = -i g \int_{0}^{\tau} A(t) e^{i\Delta_b t}\, dt.
\end{equation}
Evaluating this integral yields the closed-form expression:
\begin{equation}
\begin{aligned}
    \langle b \rangle_\tau = -ig \Bigg[ & C_+ \frac{e^{\left(-\frac{\alpha}{2} + \Omega + i\Delta_b\right)\tau} - 1}{-\frac{\alpha}{2} + \Omega + i\Delta_b} + C_- \frac{e^{\left(-\frac{\alpha}{2} - \Omega + i\Delta_b\right)\tau} - 1}{-\frac{\alpha}{2} - \Omega + i\Delta_b} \\
    & + C_0 \frac{e^{i\Delta_b \tau} - 1}{i\Delta_b} + \left(\frac{C_1 - iC_2}{2}\right) \frac{e^{i(\Delta_b + 2\omega_d)\tau} - 1}{i(\Delta_b + 2\omega_d)} \\
    & + \left(\frac{C_1 + iC_2}{2}\right) \frac{e^{i(\Delta_b - 2\omega_d)\tau} - 1}{i(\Delta_b - 2\omega_d)} \Bigg].
\end{aligned}
\label{eq:b_exact_solution}
\end{equation}
According to Eq.~\eqref{eq:ergotropy_final}, the battery ergotropy is governed by this coherent displacement:
\begin{equation}
    \mathcal{E}_B(\tau) = \omega_b |\langle b \rangle_\tau|^2.
\end{equation}

\begin{figure}[t]
    \centering
    \includegraphics[width=\columnwidth]{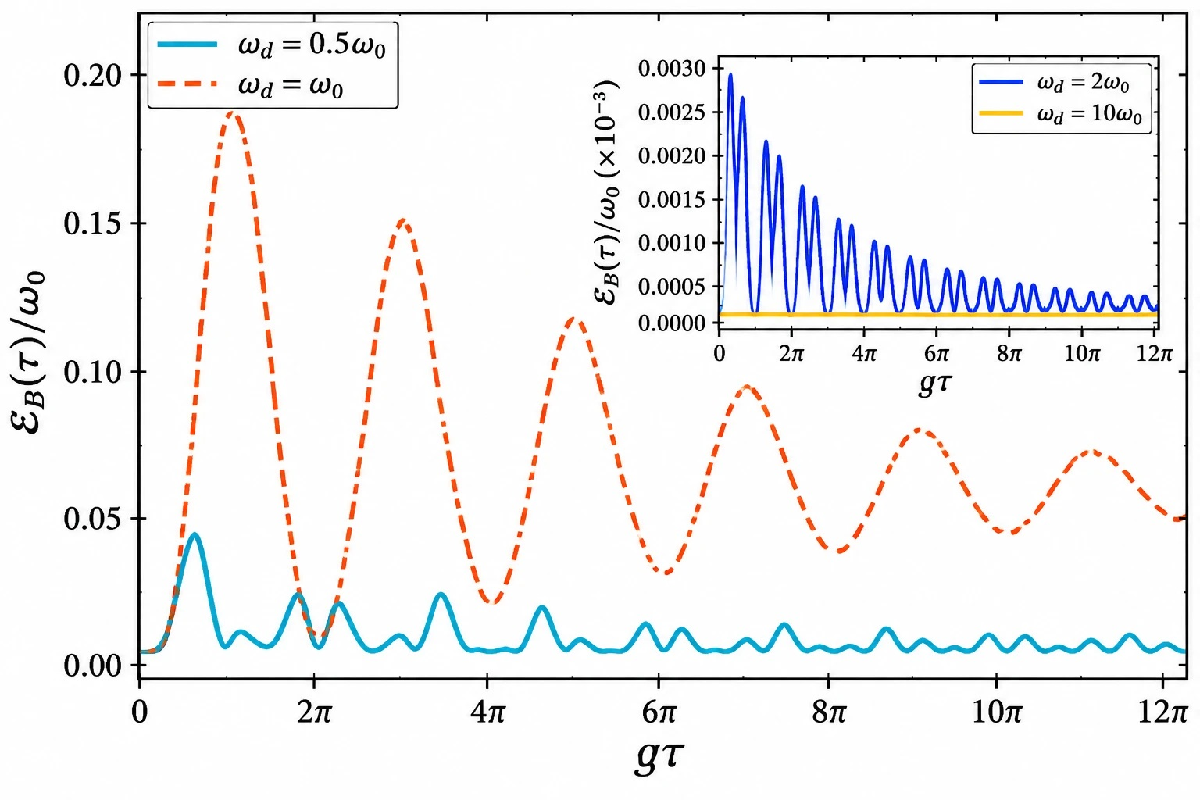}
    \caption{\label{fig:ergotropy_evolution}
    Dimensionless ergotropy $\mathcal{E}_B(\tau)/\omega_0$ versus scaled charging time $g\tau$ for various driving frequencies $\omega_d$ in the underdamped regime ($\gamma = 0.05\omega_0 < g = 0.2\omega_0$). Main panel: resonant and near-resonant driving ($\omega_d = \omega_0, 0.5\omega_0$). Inset: off-resonant high-frequency driving ($\omega_d = 2\omega_0, 10\omega_0$). Other parameters are set to $\omega_a = \omega_b = \omega_0$ and $\varepsilon_0 = 0.1\omega_0$.
    }
\end{figure}

\begin{figure}[t]
    \centering
    \includegraphics[width=\columnwidth]{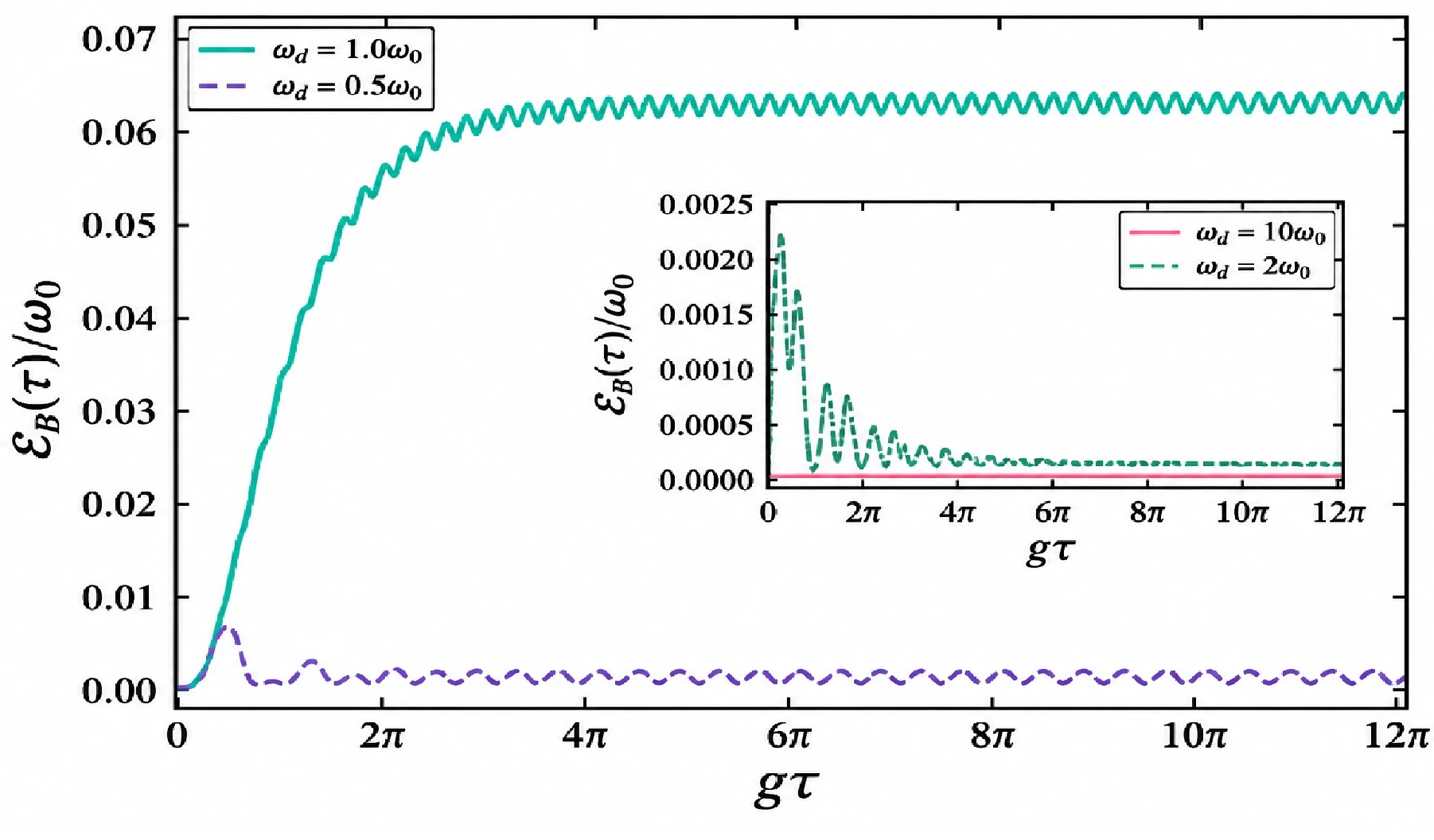}
    \caption{\label{fig:overdamped_ergotropy}
    Charging dynamics of the battery ergotropy $\mathcal{E}_B(\tau)/\omega_0$ in the overdamped regime ($\gamma = \omega_0 \gg g = 0.2\omega_0$). Main panel: transition to an asymptotic steady state for $\omega_d = \omega_0$ and $0.5\omega_0$, where strong environmental decay suppresses reversible energy transfer. Inset: suppression of extractable work under high-frequency driving ($\omega_d = 2\omega_0, 10\omega_0$). System parameters are identical to Fig.~\ref{fig:ergotropy_evolution} except for $\gamma$.
    }
\end{figure}

In Fig.~\ref{fig:ergotropy_evolution}, we illustrate the temporal evolution of the normalized ergotropy $\mathcal{E}_B(\tau)/\omega_0$ as a function of the dimensionless charging time $g\tau$ under resonant conditions ($\omega_a = \omega_b = \omega_0$). We first examine the underdamped regime, where $\gamma = 0.05\omega_0 < g = 0.2\omega_0$. Here, the coherent inter-mode coupling rate dominates over environmental loss, enabling characteristic Rabi-like energy exchange between charger and battery.

As shown in the main panel of Fig.~\ref{fig:ergotropy_evolution}, optimal work extraction occurs under exact resonance ($\omega_d = \omega_0$), establishing a pronounced peak ergotropy $\mathcal{E}_B(\tau)/\omega_0 \approx 0.20$. The subsequent damped oscillatory pattern reflects the coherent circulation of energy competing with dissipation. Conversely, detuned sub-resonant driving ($\omega_d = 0.5\omega_0$) induces phase mismatch, resulting in irregular energy exchange and a lower charging maximum. Driving well above resonance ($\omega_d = 2\omega_0, 10\omega_0$), depicted in the inset, hinders energy transfer, causing the ergotropy to drop by several orders of magnitude.

In Fig.~\ref{fig:overdamped_ergotropy}, we examine the opposite regime of strong environmental dissipation ($\gamma = \omega_0 \gg g$). When dissipation dominates over coherent coupling, reversible energy oscillations are suppressed. Under resonant driving ($\omega_d = \omega_0$), the ergotropy displays no oscillatory behavior; after an initial transient period, it settles monotonically into a stationary plateau at $\mathcal{E}_B/\omega_0 \approx 0.063$. This suppression embodies a continuous quantum Zeno effect: rapid decay into the thermal bath effectively monitors and drains coherence from mode $A$ before it can be transferred to mode $B$. For off-resonant and high-frequency driving ($\omega_d = 2\omega_0, 10\omega_0$, inset), the driving field averages to zero on the characteristic relaxation timescale $\tau_{\mathrm{rel}} \sim 1/\gamma$, quenching the charging process almost entirely.

\section{Counter-Diabatic Driving in Closed and Open Quantum Systems}
\label{sec:CD_dynamics}

\subsection{Closed-system adiabatic dynamics and counter-diabatic driving}
\label{subsec:closed_CD}

We first briefly outline the canonical formulation of counter-diabatic (CD) driving in closed quantum systems~\cite{d17, d18}. Consider a time-dependent reference Hamiltonian $H_0(t)$ with instantaneous eigenstates $|n(t)\rangle$ satisfying $H_0(t)|n(t)\rangle = E_n(t)|n(t)\rangle$. Rapid modulation of the Hamiltonian parameters induces non-adiabatic transitions among these instantaneous eigenstates, precluding faithful tracking of the target state. To suppress these excitations, an auxiliary counter-diabatic Hamiltonian $H_{\mathrm{CD}}(t)$ is introduced, leading to the full driving Hamiltonian:
\begin{equation}
    H(t) = H_0(t) + H_{\mathrm{CD}}(t),
    \label{eq:total_H_CD}
\end{equation}
where $H_{\mathrm{CD}}(t)$ cancels all diabatic transitions and takes the explicit form:
\begin{equation}
    H_{\mathrm{CD}}(t) = i\hbar \sum_n \Big[ |\partial_t n(t)\rangle\langle n(t)| - \langle n(t)|\partial_t n(t)\rangle |n(t)\rangle\langle n(t)| \Big].
    \label{eq:H_CD_closed}
\end{equation}
Under the unitary evolution governed by $H(t)$, the state tracks the instantaneous eigenstates of $H_0(t)$ without diabatic excitations, irrespective of the driving speed.

\subsection{Generalization to open quantum systems in Liouville space}
\label{subsec:open_CD_formalism}

In physical quantum batteries, coupling to the surrounding environment is unavoidable. Transitionless tracking in open quantum architectures can be systematically formulated within Liouville space~\cite{d42, d45}. The reduced density operator $\rho(t)$ evolves according to the GKSL master equation~\cite{d35, d36, d37}:
\begin{equation}
    \dot{\rho}(t) = \mathcal{L}(t)[\rho(t)],
    \label{eq:master_liouville}
\end{equation}
where $\mathcal{L}(t)$ denotes the Liouvillian superoperator. 

To track the instantaneous steady state under dissipation, we transform the Liouvillian into its Jordan canonical form (JCF) via a time-dependent similarity transformation $\hat{\mathcal{O}}(t)$ in Liouville space~\cite{d42}:
\begin{equation}
    \hat{\mathcal{O}}^{-1}(t)\mathcal{L}(t)\hat{\mathcal{O}}(t) = \operatorname{diag}\big(J_1(t), J_2(t), \dots, J_N(t)\big).
    \label{eq:JCF_liouvillian}
\end{equation}
Defining the instantaneous adiabatic frame via $\rho'(t) = \hat{\mathcal{O}}^{-1}(t)[\rho(t)]$, the equation of motion transforms to:
\begin{equation}
    \dot{\rho}'(t) = \hat{\mathcal{O}}^{-1}(t)\mathcal{L}(t)\hat{\mathcal{O}}(t)[\rho'(t)] - \hat{\mathcal{O}}^{-1}(t)\dot{\hat{\mathcal{O}}}(t)[\rho'(t)].
    \label{eq:adiabatic_frame_evolution}
\end{equation}
Transitions between distinct Jordan blocks induced by the inertial gauge potential $-\hat{\mathcal{O}}^{-1}(t)\dot{\hat{\mathcal{O}}}(t)$ can be suppressed by adding an auxiliary CD Liouvillian $\mathcal{L}_{\mathrm{CD}}(t)$ such that $\hat{\mathcal{O}}^{-1}(t)\mathcal{L}_{\mathrm{CD}}(t)\hat{\mathcal{O}}(t) = \hat{\mathcal{O}}^{-1}(t)\dot{\hat{\mathcal{O}}}(t)$.

\subsection{Local counter-diabatic control for the charger}
\label{subsec:local_CD_charger}

We now apply this open-system framework to the composite battery architecture, where the quantum battery ($B$) is charged via a driven-dissipative harmonic charger ($A$). In this layout, the charger is coupled directly to the external control field $\varepsilon(t)$ and the thermal bath. Conversely, the battery remains isolated from both the environment and the external driving field, exchanging energy exclusively through the coherent inter-mode coupling $H_{\text{int}} = g(a^\dagger b + a b^\dagger)$. As a result, non-adiabatic transitions occur entirely within subsystem $A$, enabling the implementation of the auxiliary CD control strictly locally on the charger mode.

In the rotating frame with respect to the drive frequency $\omega_d$, the bare Hamiltonian of the driven charger is
\begin{equation}
H_C(t) = \Delta_a a^\dagger a - i \left( \varepsilon(t) a^\dagger - \varepsilon^*(t) a \right),
\label{eq:charger_hamiltonian}
\end{equation}
representing the standard form of a coherent drive in the interaction picture~\cite{d45}. For this driven Gaussian charger mode, the similarity superoperator $\hat{\mathcal{O}}(t)$ is represented by the displacement superoperator $\hat{\mathcal{D}}(\bar{\alpha}(t))$, defined by its action on the density matrix as
\begin{equation}
    \hat{\mathcal{D}}(\bar{\alpha}(t))[\rho] \equiv D(\bar{\alpha}(t))\rho D^{\dagger}(\bar{\alpha}(t)),
    \label{eq:displacement_superop}
\end{equation}
where $D(\bar{\alpha}) = \exp(\bar{\alpha} a^\dagger - \bar{\alpha}^* a)$ is the standard unitary displacement operator. Under this transformation, the field operator shifts according to $D^\dagger(\bar{\alpha}) a D(\bar{\alpha}) = a + \bar{\alpha}$. Setting the instantaneous displacement amplitude to track the steady state of the driven-dissipative cavity yields
\begin{equation}
    \bar{\alpha}(t) = \frac{i\varepsilon(t)}{\Delta_a - i\gamma/2},
    \label{eq:alpha_steady_state}
\end{equation}
which is fixed by the condition $\dot{\bar{\alpha}}_{\mathrm{ss}}(t) = 0$. Moving into the displaced thermal frame via $D[\bar{\alpha}(t)]$ removes the classical drift terms from the first-order moment equations. An explicit derivation of $\bar{\alpha}(t)$, the vanishing of the thermal noise contributions in the displaced frame, and the resulting gauge potential is provided in Appendix~\ref{app:appendix_F}. In this frame, the transformed Liouvillian $\mathcal{L}_{\mathcal{J}} \equiv \hat{\mathcal{D}}^{-1}(\bar{\alpha}(t))\mathcal{L}(t)\hat{\mathcal{D}}(\bar{\alpha}(t))$ becomes time-independent.

Defining the moving-frame density operator as $\rho'(t) = \hat{\mathcal{D}}^{-1}(\bar{\alpha}(t))[\rho(t)] = D^\dagger(\bar{\alpha}(t))\rho(t)D(\bar{\alpha}(t))$, the transformed equation of motion reads
\begin{equation}
    \dot{\rho}'(t) = -i \big[ i\dot{D}^\dagger(\bar{\alpha})D(\bar{\alpha}),\, \rho'(t) \big] + \mathcal{L}_{\mathcal{J}}[\rho'(t)],
    \label{eq:displaced_frame_eom}
\end{equation}
which identifies the non-adiabatic inertial term as $H_{\mathrm{i}}(t) = i\dot{D}^\dagger(\bar{\alpha})D(\bar{\alpha}) = -i D^\dagger(\bar{\alpha})\dot{D}(\bar{\alpha})$. Introducing the CD Liouvillian $\mathcal{L}_{\mathrm{CD}}[\rho] = -i[H_{\mathrm{CD}}(t), \rho]$ cancels this inertial contribution when mapped back to the laboratory frame. Differentiating Eq.~\eqref{eq:alpha_steady_state} with respect to time yields
\begin{equation}
    \dot{\bar{\alpha}}(t) = \frac{i\dot{\varepsilon}(t)}{\Delta_a - i\gamma/2},
    \label{eq:alpha_dot}
\end{equation}
which provides the explicit local CD Hamiltonian:
\begin{equation}
    H_{\mathrm{CD}}(t) = i \dot{D}(\bar{\alpha}) D^\dagger(\bar{\alpha}) = -\frac{\dot{\varepsilon}(t)}{\Delta_a - i\gamma/2} a^\dagger + \text{h.c.}
    \label{eq:H_CD_explicit}
\end{equation}
Accordingly, counter-diabatic driving in this open Gaussian system is implemented entirely through an auxiliary local coherent field applied directly to the charger.

This local control ensures that the charger mode tracks its instantaneous steady-state manifold, accelerating energy transfer to the battery while mitigating non-adiabatic lag. As shown in Eq.~\eqref{eq:H_CD_explicit}, the CD protocol modifies the effective coherent drive field. In addition to the bare pulse $\varepsilon(t)$, the total effective drive acquires a derivative correction:
\begin{equation}
    \varepsilon_{\mathrm{CD}}(t) = \varepsilon(t) - i\frac{\dot{\varepsilon}(t)}{\Delta_a - i\gamma/2}.
    \label{eq:eps_CD_field}
\end{equation}

\begin{figure}[!t]
    \centering
    \includegraphics[width=\columnwidth]{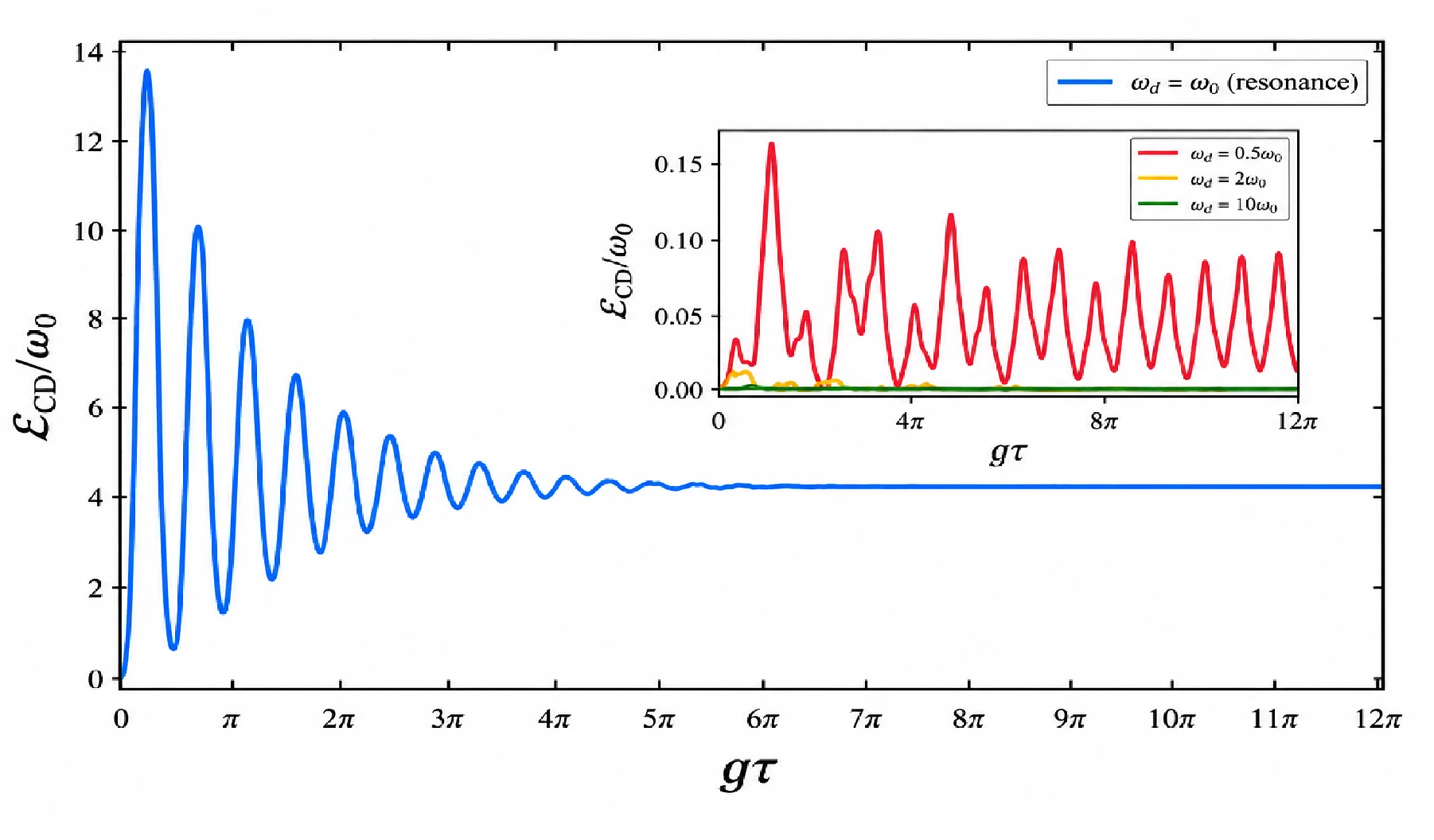}
    \caption{
       Ergotropy evolution of the quantum battery under counter-diabatic driving in the underdamped regime ($\gamma = 0.05\,\omega_0 < g = 0.2\,\omega_0$). The normalized ergotropy $\mathcal{E}_{\mathrm{CD}}/\omega_0$ is shown as a function of the dimensionless charging time $g\tau$ for driving frequencies $\omega_d = 0.5\omega_0, \omega_0, 2\omega_0,$ and $10\omega_0$. Other parameters are $\omega_a = \omega_b = \omega_0$ and $\varepsilon_0 = 0.1\,\omega_0$.
    }
    \label{fig:ergotropy-evolution}
\end{figure}

\begin{figure}[t]
    \centering
     \includegraphics[width=\columnwidth]{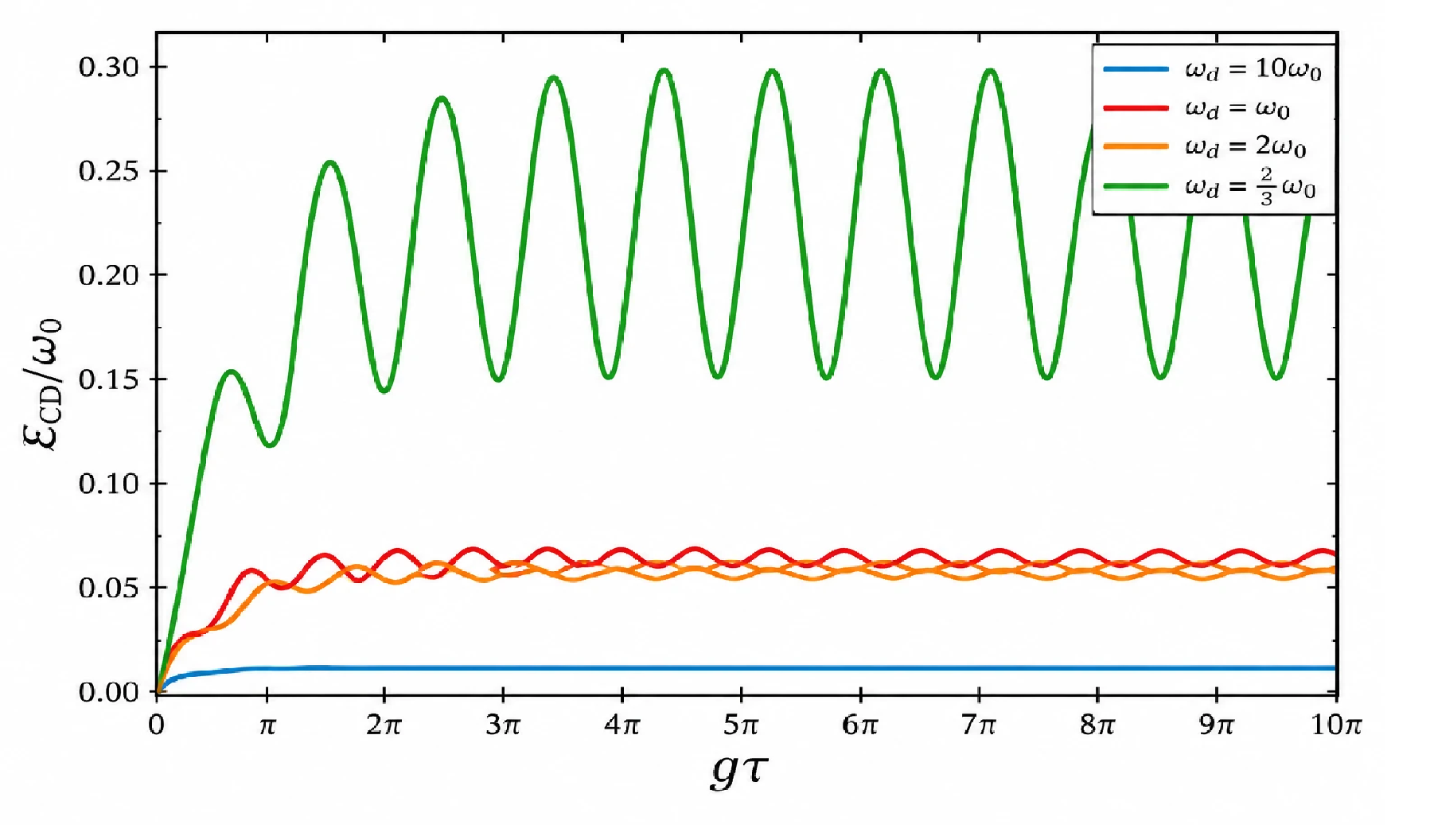}
    \caption{
       Ergotropy evolution of the quantum battery under counter-diabatic driving in the overdamped regime ($\gamma = \omega_0 \gg g = 0.2\,\omega_0$). The scaled ergotropy $\mathcal{E}_{\mathrm{CD}}/\omega_0$ is plotted against $g\tau$ for the same four driving frequencies $\omega_d$. Parameters are identical to Fig.~\ref{fig:ergotropy-evolution} except for the dissipation rate $\gamma = \omega_0$.
    }
    \label{fig:stored-energy-cd}
\end{figure}
Counter-diabatic driving governs the extractable work exclusively through the acquired coherent displacement amplitude. While the relative phase of $\langle b(\tau)\rangle_{\mathrm{CD}}$ determines the interference among transient modes, only its magnitude enters the ergotropy.

Figures~\ref{fig:ergotropy-evolution} and~\ref{fig:stored-energy-cd} illustrate the charging behavior of the battery via the scaled ergotropy $\mathcal{E}_B(\tau)/\omega_0$ versus $g\tau$ across different driving frequencies $\omega_d$. In both dissipation regimes, the extractable work exhibits pronounced frequency selectivity: driving frequencies far from resonance lead to marginal energy transfer, whereas near-resonant driving facilitates substantial ergotropy accumulation.

In the underdamped regime (Fig.~\ref{fig:ergotropy-evolution}), the dynamics is governed by coherent energy exchange between charger and battery, yielding well-resolved Rabi oscillations and reaching maximal ergotropy on short timescales. In the overdamped regime (Fig.~\ref{fig:stored-energy-cd}), strong environmental dissipation dampens the oscillatory behavior; nevertheless, the frequency dependence remains qualitatively identical. Transitionless tracking thus protects the charging process against non-adiabatic dispersion, preserving the advantages of fast driving in lossy open quantum systems.

\subsection{Numerical Results and Discussion}
\label{subsec:numerical_results_discussion}

We now quantitatively compare the charging dynamics generated by normal driving (ND) and counter-diabatic driving (CD). The comparison is conducted for the same battery--charger model using identical physical parameters; only the driving protocol is varied. The primary figure of merit is the ergotropy $\mathcal{E}_{B}(\tau)$, quantifying the maximal extractable work via cyclic unitaries.

The weakly dissipative case corresponds to $\gamma = 0.05\,\omega_{0}$ (underdamped regime). The corresponding ND and CD profiles are shown in Figs.~\ref{fig:ergotropy_evolution} and~\ref{fig:ergotropy-evolution}, respectively. The strongly dissipative case is defined by setting $\gamma = \omega_{0}$ with $g = 0.2\,\omega_{0}$, such that
\begin{equation}
    \frac{\gamma}{g} = 5,
\end{equation}
meaning that environmental decay acts on a timescale considerably shorter than the coherent coupling rate. This overdamped regime is shown for ND and CD in Figs.~\ref{fig:overdamped_ergotropy} and~\ref{fig:stored-energy-cd}, respectively.

Under normal driving in the underdamped regime, Fig.~\ref{fig:ergotropy_evolution} exhibits oscillatory coherent energy exchange between the charger and the battery. The amplitude of these oscillations depends on $\omega_{d}$, reflecting the competition among the driving timescale, coupling $g$, and loss rate $\gamma$. While the battery accumulates non-zero work, the maximum scaled ergotropy remains around $0.2\,\omega_{0}$ even under optimal resonance.

In contrast, the CD protocol shown in Fig.~\ref{fig:ergotropy-evolution} demonstrates a substantial enhancement. The ergotropy accumulates on a shorter timescale and reaches values close to $14\,\omega_{0}$ under resonant driving ($\omega_d = \omega_0$). Compared to the bare ND baseline, this constitutes an improvement by approximately a factor of
\begin{equation}
    \frac{14}{0.2} \simeq 70,
\end{equation}
achieving an increase of nearly two orders of magnitude in extractable energy. Because the CD protocol supplements the drive with the time-derivative term proportional to $\dot{\varepsilon}(t)$ according to Eq.~\eqref{eq:eps_CD_field}, the modified field actively compensates for the non-adiabatic phase lag induced by cavity dissipation and finite response time. This compensation enhances both the maximum ergotropy and the instantaneous charging rate.

The detrimental role of dissipation is most pronounced in the overdamped regime ($\gamma/g = 5$). As shown in Fig.~\ref{fig:overdamped_ergotropy}, normal driving leads to heavily suppressed charging. Under resonance ($\omega_{d} = \omega_{0}$), the ergotropy quickly saturates to a stationary plateau of $\mathcal{E}_{B}/\omega_{0} \simeq 0.063$, while oscillatory features vanish. For detuned driving ($\omega_{d} = 0.5\,\omega_{0}$), the maximum ergotropy drops below $0.01\,\omega_{0}$, and for high-frequency driving ($\omega_{d} = 2\,\omega_{0}, 10\,\omega_{0}$), the accumulated energy is negligible.

Figure~\ref{fig:stored-energy-cd} shows that the CD protocol mitigates this dissipation-induced suppression under identical overdamped parameters. While CD control does not eliminate physical decay channels, it tailors the coherent drive field so that energy transfer remains robust against dissipative lag. The resulting dynamics preserves clear frequency selectivity and achieves higher ergotropy than the ND baseline across all considered values of $\omega_{d}$. This improvement is particularly pronounced in non-adiabatic driving regimes, demonstrating that counter-diabatic control acts as an effective shield against environmental degradation during fast charging.
\subsection{Steady states and adiabatic timescale}
\label{subsec:steady_states_adiabatic_timescale}

In the displaced adiabatic frame, the dynamics of the driven open charger is governed by a GKSL-type Liouvillian of the form
\begin{equation}
    \mathcal{L}_{\mathcal{J}}\rho = -i[\Delta_a a^\dagger a, \rho] + \gamma (N_b + 1)\mathcal{D}[a]\rho + \gamma N_b \mathcal{D}[a^\dagger]\rho,
    \label{eq:LJ}
\end{equation}
where the Lindblad dissipator is defined as
\begin{equation}
    \mathcal{D}[L]\rho = L\rho L^\dagger - \frac{1}{2}\{L^\dagger L, \rho\},
\end{equation}
$\Delta_a$ is the cavity detuning, $\gamma$ is the dissipation rate of the charger mode, and the average thermal occupation number $N_b$ accounts for the finite temperature of the reservoir. Compared to the zero-temperature scenario, the thermal excitation channel $\mathcal{D}[a^\dagger]$ modifies the matrix representation of $\mathcal{L}_{\mathcal{J}}$ in the Fock basis while preserving its block-tridiagonal structure across invariant subspaces. Its explicit action on the basis operators $|m\rangle\langle n|$ reads
\begin{align}
    \mathcal{L}_{\mathcal{J}}|m\rangle\langle n| ={}& \biggl[ -i\Delta_a(m - n) - \frac{\gamma}{2}(2N_b + 1)(m + n) - \gamma N_b \biggr]|m\rangle\langle n| \nonumber\\
    & + \gamma (N_b + 1)\sqrt{mn}\,|m - 1\rangle\langle n - 1| \nonumber\\
    & + \gamma N_b\sqrt{(m + 1)(n + 1)}\,|m + 1\rangle\langle n + 1|.
    \label{eq:LJ-action}
\end{align}

Consequently, for each fixed index $j = m - n$, the Liouvillian leaves invariant the operator subspace
\begin{equation}
    \mathcal{H}_{j} = \mathrm{span}\!\left\{ |k\rangle\langle k + j| \right\}_{k=0}^{\infty}.
    \label{eq:Hj}
\end{equation}
The right eigenoperators of $\mathcal{L}_{\mathcal{J}}$ can therefore be expanded as
\begin{equation}
    \rho_j = \sum_{k=0}^{\infty} c_k^{(j)} |k\rangle\langle k + j|.
    \label{eq:rhoj-expansion}
\end{equation}
Substituting this ansatz into the eigenvalue equation $\mathcal{L}_{\mathcal{J}}\rho_j = e \rho_j$ yields a three-term recurrence relation for the coefficients $c_k^{(j)}$. Solving this via the generating-function method yields the complete eigenvalue spectrum:
\begin{equation}
    e_{j,k} = i\Delta_a j - \gamma\left(\frac{|j|}{2} + k\right), \qquad j \in \mathbb{Z}, \quad k = 0, 1, 2, \dots.
    \label{eq:ejk}
\end{equation}
The detailed step-by-step derivation is provided in Appendix~\ref{app:biorthogonal_spectra}. 
Remarkably, the thermal occupation number $N_b$ enters the structure of the eigenoperators and the steady-state distribution, but leaves the functional form of the eigenvalue spectrum completely invariant. The unique zero-eigenvalue steady state corresponds to $j = 0$ and $k = 0$ ($e_{0,0} = 0$), which is the thermal state
\begin{equation}
    \rho_{\mathrm{ss}}^{(\mathcal{J})} = \frac{1}{N_b + 1}\sum_{k=0}^{\infty} \left(\frac{N_b}{N_b + 1}\right)^k |k\rangle\langle k|.
    \label{eq:rho-ss-J}
\end{equation}
Transforming back to the laboratory frame, the instantaneous steady state is given by the displaced thermal state:
\begin{equation}
    \rho_{\mathrm{ss}}(t) = D(\bar{\alpha}(t))\, \rho_{\mathrm{ss}}^{(\mathcal{J})}\, D^\dagger(\bar{\alpha}(t)).
    \label{eq:rho-ss-lab}
\end{equation}

The open-system adiabatic criterion in Liouville space can be formulated as \cite{d46}
\begin{equation}
    \max_{0 < t < t_f} \left| \left\langle \tilde{\rho}_{j',k'}(t),\, \dot{\rho}_{j,k}(t) \right\rangle \right| \ll \left| e_{j,k} - e_{j',k'} \right|,
    \label{eq:adiabatic-condition}
\end{equation}
where $\langle u, v \rangle = \mathrm{Tr}(u^\dagger v)$ denotes the Hilbert-Schmidt inner product, $\rho_{j,k}(t)$ are the instantaneous right eigenoperators of $\mathcal{L}(t)$, and $\tilde{\rho}_{j,k}(t)$ are the corresponding left eigenoperators satisfying the biorthogonality relation. Parameterizing the evolution by the normalized time $s = t/t_f$ such that $|\dot{\rho}_{j,k}| \sim t_f^{-1}$, this yields the sufficient adiabatic condition:
\begin{equation}
    t_f \gg \frac{1}{\min_{(j,k) \neq (j',k')} |e_{j,k} - e_{j',k'}|} = \frac{1}{\min\!\left(\sqrt{\Delta_a^2 + \gamma^2/4},\, \gamma\right)}.
    \label{eq:tf-gap}
\end{equation}

Because the Liouvillian spectrum is independent of $N_b$, the fundamental adiabatic timescale is unaffected by bath thermal excitations at the level of the dissipative spectral gap:
\begin{equation}
    t_f \gg \left[ \min\!\left(\sqrt{\Delta_a^2 + \gamma^2/4},\, \gamma\right) \right]^{-1}.
    \label{eq:adiabatic-scale}
\end{equation}
On resonance ($\Delta_a = 0$), this condition reduces directly to
\begin{equation}
    t_f \gg \frac{2}{\gamma}.
    \label{eq:resonant-adiabatic}
\end{equation}

For the smooth driving pulse
\begin{equation}
    \varepsilon(t) = \varepsilon_0 \sin^2(\omega_d t), \qquad \omega_d = \frac{\pi}{2 t_f},
    \label{eq:smooth-protocol}
\end{equation}
in the experimentally ubiquitous regime where $\Delta_a \lesssim \gamma$, the adiabatic timescale is essentially dictated by the dissipation rate: $t_f \gg 2\gamma^{-1}$. In the underdamped regime ($\gamma \ll \omega_0$), this requires exceptionally long protocol durations that are impractical for fast battery charging and leave the system vulnerable to environmental decoherence. Consequently, bare adiabatic charging is effectively unviable in the weakly damped domain. 

In sharp contrast, counter-diabatic driving completely suppresses transient diabatic excitations across the instantaneous Liouvillian eigenspaces, enabling the system to strictly track the target trajectory within an arbitrary fast protocol duration $t_f$, irrespective of whether the dynamics is underdamped or overdamped. This establishes shortcut-to-adiabaticity as an indispensable mechanism for high-power, high-capacity quantum battery charging beyond the intrinsic limitations of standard adiabatic protocols.

\section{Energetic Cost and Thermodynamic Viability of Counter-Diabatic Charging}
\label{sec:energetic_cost}

To establish whether the proposed counter-diabatic (CD) charging protocol confers a genuine thermodynamic advantage, the ergotropy stored in the quantum battery must be rigorously benchmarked against the total energetic investment required to implement the driving protocol. In open quantum systems, accelerating the charging dynamics inevitably incurs an auxiliary-field control cost, while simultaneously suppressing cumulative environmental dissipation. In this section, we formulate a comprehensive energy-overhead assessment, evaluate the operational control fluence, and demonstrate the net thermodynamic advantage of the CD driving strategy across both underdamped and overdamped regimes.

\subsection{Field Fluence and Thermodynamic Figure of Merit}
\label{subsec:fluence_econ_index}

Following optimal quantum control theory, the energetic cost of an external control field is characterized by its temporal fluence, defined as the time-integrated intensity of the driving field. The total charging cost $C_{\mathrm{total}}(\tau)$ over the protocol duration $\tau$ accounts for both the primary drive fluence $C_{\mathrm{drive}}(\tau)$ and the counter-diabatic auxiliary-field fluence $C_{\mathrm{CD}}(\tau)$~\cite{Werschnik2007}:
\begin{equation}
    C_{\mathrm{total}}(\tau) = C_{\mathrm{drive}}(\tau) + \xi C_{\mathrm{CD}}(\tau),
    \label{eq:total_cost}
\end{equation}
where $\xi > 0$ is a dimensionless engineering weight quantifying the relative experimental overhead and power-delivery constraints of the auxiliary control channel. Rather than fixing an arbitrary fiducial value for $\xi$, we perform a sensitivity analysis across the realistic experimental window $5 \le \xi \le 20$, verifying that the thermodynamic conclusions remain robust throughout this parameter range.

For the smooth sine-squared envelope $\varepsilon(t) = \varepsilon_0 \sin^2(\omega_d t)$ with protocol duration $\tau \equiv t_f = \pi / (2\omega_d)$, the boundary conditions $\varepsilon(0) = \dot{\varepsilon}(0) = \dot{\varepsilon}(t_f) = 0$ ensure smooth switching without non-adiabatic boundary transients. The primary drive fluence over the duration $\tau$ is evaluated as
\begin{equation}
    C_{\mathrm{drive}}(\tau) = \int_0^{\tau} |\varepsilon(t)|^2 \, dt = \frac{3}{8}\varepsilon_0^2 \tau.
    \label{eq:C_drive}
\end{equation}
Compared to a sudden square pulse of identical amplitude ($C_{\mathrm{sq}} = \varepsilon_0^2 \tau / 2$), this smooth pulse profile achieves a $62.5\%$ reduction in drive fluence, significantly lowering baseline energy expenditures.

The auxiliary CD Hamiltonian derived in Sec.~\ref{sec:CD_dynamics} couples locally to the charger with an effective control amplitude $\chi(t) = \dot{\varepsilon}(t)/(\Delta_a - i\gamma/2)$. The associated CD field fluence is given by
\begin{equation}
    C_{\mathrm{CD}}(\tau) = \frac{1}{\Delta_a^2 + \gamma^2/4} \int_0^{\tau} |\dot{\varepsilon}(t)|^2 \, dt = \frac{\varepsilon_0^2 \omega_d^2\, \tau}{2\left(\Delta_a^2 + \dfrac{\gamma^2}{4}\right)},
    \label{eq:C_CD}
\end{equation}
where the factor of $1/2$ stems from integrating $\sin^2(2\omega_d t)$ over exactly half of its oscillation period, $\tau = \pi/(2\omega_d)$.

To quantify the net operational performance of the charging process, we introduce the dimensionless net charging efficiency $\eta_{\mathrm{net}}$ as
\begin{equation}
    \eta_{\mathrm{net}} \equiv \frac{\mathcal{E}_B(\tau)}{C_{\mathrm{total}}(\tau) + Q_{\mathrm{diss}}(\tau)},
    \label{eq:eta_net}
\end{equation}
where $\mathcal{E}_B(\tau)$ is the extractable ergotropy stored in the quantum battery, and
\begin{equation}
    Q_{\mathrm{diss}}(\tau) = \int_0^\tau \operatorname{Tr}\left[\mathcal{D}(\rho_t) H_0\right] dt
    \label{eq:Q_diss}
\end{equation}
quantifies the cumulative heat dissipated into the reservoir during charging, with the sign convention that $Q_{\mathrm{diss}} > 0$ denotes net energy transfer from the system to the environment, consistent with the standard framework of quantum thermodynamics~\cite{d3,Breuer2002}. In Eq.~\eqref{eq:eta_net}, the denominator encompasses the comprehensive energetic footprint by unifying external operational control expenditures with irreversible bath losses into a single thermodynamic figure of merit. 

Crucially, the order-of-magnitude superiority of the CD driving protocol over unassisted charging is preserved across the entire empirical overhead range $5 \le \xi \le 20$, confirming that the observed advantage is not an artifact of idealized control assumptions.

\begin{table*}[t]
\centering
\caption{Thermodynamic performance comparison between unassisted driving (ND) and counter-diabatic driving (CD) under on-resonance conditions ($\Delta_a = 0$, $\omega_d = \omega_0$).}
\label{tab:thermodynamic_comparison}
\begin{ruledtabular}
\begin{tabular}{lcccc}
Regime & Protocol & Ergotropy $\mathcal{E}_B/\omega_0$ & Dissipated Heat $Q_{\mathrm{diss}}$ & Operational Viability ($\eta_{\mathrm{net}}$) \\
\hline
\textbf{Underdamped} ($\gamma = 0.05\,\omega_0$) & ND & $\approx 0.20$ & Finite & Baseline \\
& CD & $\approx 14.0$ ($\sim 70\times$) & $Q_{\mathrm{diss}} \to 0$ & Enhanced ($\sim 10^2\times$) \\
\hline
\textbf{Overdamped} ($\gamma = \omega_0$) & ND & $\to 0$ (decayed) & High & Unviable ($\eta_{\mathrm{net}} \to 0$) \\
& CD & $> 0$ (sustained) & Suppressed & Restored ($\eta_{\mathrm{net}} > 0$) \\
\end{tabular}
\end{ruledtabular}
\end{table*}

\subsection{Thermodynamic Performance in Underdamped and Overdamped Regimes}
\label{subsec:regimes_performance}

\subsubsection{Underdamped Regime ($\gamma = 0.05\,\omega_0 < g = 0.2\,\omega_0$)}
In the weak-dissipation regime, the unassisted normal driving (ND) protocol at resonance achieves a maximum extractable ergotropy of only $\mathcal{E}_B^{\mathrm{ND}} \approx 0.20\,\omega_0$ (see baseline in Fig.~\ref{fig:ergotropy_evolution}), which rapidly drops for off-resonant drive frequencies ($\omega_d \ge 2\,\omega_0$). The prolonged protocol duration required by bare driving inherently leads to substantial accumulated thermal dissipation $Q_{\mathrm{diss}}$.

In sharp contrast, the CD driving protocol enforces rapid coherent state transfer on a timescale $t_f \ll \gamma^{-1}$, drastically shortening the exposure to the reservoir and effectively suppressing thermal losses such that $Q_{\mathrm{diss}} \to 0$. As evidenced by the resonant ergotropy peak in Fig.~\ref{fig:ergotropy-evolution}, the extractable ergotropy reaches $\mathcal{E}_B^{\mathrm{CD}} \approx 14.0\,\omega_0$, yielding a 70-fold enhancement relative to the ND baseline. Because the invested control fluence remains modest ($C_{\mathrm{total}} \simeq 25.1\,\omega_0$) owing to the moderate driving amplitude $\varepsilon_0 = 0.1\,\omega_0$, the net charging efficiency $\eta_{\mathrm{net}}$ is enhanced by nearly two orders of magnitude ($\sim 10^2\times$) compared to the unassisted benchmark. This significant gain persists robustly across the entire parameter window $\xi \in [5, 20]$.

\subsubsection{Overdamped Regime ($\gamma = \omega_0 \gg g = 0.2\,\omega_0$)}

In strongly dissipative environments, bare charging protocols fail catastrophically (Fig.~\ref{fig:overdamped_ergotropy}): the injected coherent energy is rapidly intercepted by bath-induced decoherence and almost entirely converted into dissipated heat ($\mathcal{E}_B^{\mathrm{ND}} \to 0$, $Q_{\mathrm{diss}} \gg \mathcal{E}_B$), rendering conventional driving thermodynamically unviable ($\eta_{\mathrm{net}}^{\mathrm{ND}} \to 0$).

Under counter-diabatic driving, the auxiliary control field suppresses non-adiabatic transitions into irreversible decay channels. By steering the system along the shortcut trajectory before severe environmental leakage takes place, $Q_{\mathrm{diss}}$ is significantly reduced, and a strictly non-zero extractable ergotropy $\mathcal{E}_B^{\mathrm{CD}} > 0$ is preserved (Fig.~\ref{fig:stored-energy-cd}). Consequently, CD driving not only optimizes the charging rate, but also acts as an indispensable stabilization mechanism that restores thermodynamic viability ($\eta_{\mathrm{net}} > 0$) in parameter regimes where standard charging is rendered completely ineffective.

In summary, these results establish that counter-diabatic driving is far from a purely mathematical prescription for high charging speeds; rather, it constitutes an energetically efficient control strategy for open quantum batteries. By drastically shortening the interaction window, curtailing cumulative bath dissipation $Q_{\mathrm{diss}}$, and exhibiting robustness against realistic experimental overheads $\xi$, the CD protocol delivers two orders of magnitude higher net efficiency in underdamped environments and successfully restores battery functionality in strongly dissipative media.

\section{Conclusion}
\label{sec:conclusion}

In this work, we have established an analytical and numerically verified framework for fast, robust charging in an open quantum battery composed of two linearly coupled continuous-variable modes. The architecture considers a driven, dissipative charger coupled to a Markovian reservoir, which indirectly transfers energy to an uncoupled battery mode that remains shielded from direct environmental decoherence throughout the operation. By formulating a local counter-diabatic (CD) driving scheme within a Lindblad master-equation description, we derived an explicit auxiliary field, 
$\varepsilon_{\mathrm{CD}}(t) = \varepsilon(t) - i\dot{\varepsilon}(t)/(\Delta_a - i\gamma/2)$, 
that dynamically offsets the inherent phase lag and non-adiabatic excitations induced by charger losses. Because the complete dynamical generator preserves Gaussianity at all times, the system evolution reduces exactly to closed sets of first- and second-order moment equations, yielding a transparent analytical expression for the stored ergotropy via the coherent battery displacement, $\mathcal{E}_B = \omega_0 |\langle b \rangle|^2$.

The resulting performance gains are quantitatively decisive across fundamentally different dissipation regimes. In the underdamped regime ($\gamma = 0.05\,\omega_0 < g = 0.2\,\omega_0$), unassisted resonant driving (ND) is severely constrained by cumulative reservoir exposure, saturating at a modest ergotropy $\mathcal{E}_B^{\mathrm{ND}} \simeq 0.20\,\omega_0$. By contrast, the CD protocol completes the charging cycle on a compressed timescale, steering the battery to $\mathcal{E}_B^{\mathrm{CD}} \simeq 14.0\,\omega_0$---a $\sim 70$-fold enhancement in extractable work---while drastically suppressing the integrated thermal leakage $Q_{\mathrm{diss}}$ via operational acceleration. In the overdamped regime ($\gamma = \omega_0 \gg g$), where standard driving fails entirely due to dissipation-induced dynamical suppression ($\mathcal{E}_B^{\mathrm{ND}}/\omega_0 \simeq 0.063$ on resonance and $<0.01$ off resonance), the CD control effectively circumvents the strong decay channel by reconfiguring the coherent drive trajectory. It thereby secures a finite, sustained ergotropy, serving as an effective control mitigation mechanism in regimes where conventional charging is thermodynamically precluded.

Crucially, the operational utility of the protocol was validated through a comprehensive energetic cost assessment. By incorporating the auxiliary control overhead into a field-fluence metric $C_{\mathrm{total}} = C_{\mathrm{drive}} + \xi C_{\mathrm{CD}}$ and scanning the empirical engineering weight over the interval $5 \le \xi \le 20$, we demonstrated that the performance advantage is not compromised by the control overhead. Supported by a smooth $\sin^2$ envelope that reduces primary fluence by $62.5\%$ compared to standard pulse shapes, the overall thermodynamic efficiency $\eta_{\mathrm{net}} = \mathcal{E}_B / (C_{\mathrm{total}} + Q_{\mathrm{diss}})$ exhibits an enhancement of nearly two orders of magnitude in the underdamped domain and restores finite efficiency in the overdamped case. Furthermore, evaluating the open-system spectral properties clarifies why bare adiabatic driving ($t_f \gg 2\gamma^{-1}$) is intrinsically incompatible with fast energy storage, directly highlighting the necessity of shortcut-to-adiabaticity protocols in dissipative quantum machinery.

Several extensions of this framework present promising avenues for further research. First, generalizing the local CD construction to multipartite or collective battery architectures (such as arrays of coupled oscillators or Tavis-Cummings topologies) could elucidate how shortcut driving scales with system size and whether collective advantages persist under finite dissipation. Second, the analytical CD trajectory derived here can serve as an informed initial ansatz for numerical optimal control or machine-learning routines, potentially minimizing auxiliary fluence under strict laboratory bandwidth constraints. Third, given the bilinear interaction and displacement-dominated dynamics, the proposed protocol is directly amenable to experimental realization in contemporary circuit quantum electrodynamics (cQED) or optomechanical setups, where drive modulation and dissipation engineering are mature capabilities. Finally, extending the energetic accounting to include quantum fluctuations of work and finite-time fluctuation relations will provide deeper insights into the non-equilibrium thermodynamics of shortcut-assisted open systems.

\allowdisplaybreaks
\raggedbottom
\appendix
\section{ Equations of Motion for First- and Second-Order Moments}
\makeatletter
\@addtoreset{equation}{section}
\makeatother
\renewcommand{\theequation}{\Alph{section}\arabic{equation}}
\label{app:equations_of_motion}
In this Appendix, we provide the step-by-step algebraic derivation of the closed dynamical equations of motion governing the first- and second-order bosonic moments in the interaction picture.
For an arbitrary time-independent system observable $X$ ($\partial X/\partial t = 0$), the quantum mechanical expectation value is defined with respect to the bipartite density operator as $\langle X \rangle(t) \equiv \operatorname{Tr}[X \tilde{\rho}_{AB}(t)]$. Differentiating with respect to time yields
\begin{equation}
    \frac{d}{dt}\langle X \rangle = \operatorname{Tr}\bigl(X \dot{\tilde{\rho}}_{AB}(t)\bigr).
    \label{eq:app_X_diff}
\end{equation}
Using the cyclic invariance of the trace, the coherent contribution generated by any interaction Hamiltonian $\tilde{H}_{\mathrm{int}}(t)$ evaluates via the identity
\begin{equation}
    -i \operatorname{Tr}\bigl(X [\tilde{H}_{\mathrm{int}}(t), \tilde{\rho}_{AB}(t)]\bigr) = -i \langle [X, \tilde{H}_{\mathrm{int}}(t)] \rangle.
    \label{eq:app_comm_identity}
\end{equation}
For the incoherent dissipative terms, the action of the Lindblad superoperator $\mathcal{D}^{[O]}[\rho] \equiv O\rho O^\dagger - \frac{1}{2}\{O^\dagger O, \rho\}$ on the trace evaluation is governed by its adjoint dual:
\begin{equation}
    \operatorname{Tr}\bigl(X \mathcal{D}_A^{[O]}[\rho]\bigr) = \operatorname{Tr}\bigl(\mathcal{D}_A^{\dagger[O]}(X)\rho\bigr) = \left\langle O^\dagger X O - \frac{1}{2}\{O^\dagger O, X\} \right\rangle.
    \label{eq:app_dissipator_dual}
\end{equation}
The microscopic interaction-picture master equation describing the charger-battery system reads
\begin{align}
    \dot{\tilde{\rho}}_{AB}(t) &= -i \left[ g(t)\left(e^{i\Delta' t}a^\dagger b + e^{-i\Delta' t}a b^\dagger\right), \tilde{\rho}_{AB}(t) \right] \nonumber \\
    &\quad - i \left[ \varepsilon(t)e^{i\Delta_a t}a^\dagger - \varepsilon^*(t)e^{-i\Delta_a t}a, \tilde{\rho}_{AB}(t) \right] \nonumber \\
    &\quad + \gamma\bigl(N_b(T) + 1\bigr)\mathcal{D}_A^{[a]}[\tilde{\rho}_{AB}(t)] \nonumber \\
    &\quad + \gamma N_b(T)\mathcal{D}_A^{[a^\dagger]}[\tilde{\rho}_{AB}(t)],
    \label{eq:app_master_eqn}
\end{align}
where $g(t) = g_0 \lambda(t)$ denotes the time-dependent inter-mode coupling.
\subsection{First-Order Moment Dynamics}
Setting $X = a$ in Eq.~\eqref{eq:app_X_diff}, we evaluate each contribution separately using canonical bosonic commutators $[a, a^\dagger] = [b, b^\dagger] = 1$ and $[a, b] = [a, b^\dagger] = 0$:
\begin{align}
    -i \left\langle \left[ a, g(t)\left(e^{i\Delta' t}a^\dagger b + e^{-i\Delta' t}a b^\dagger\right) \right] \right\rangle &= -i g(t)e^{i\Delta' t}\langle b \rangle, \label{eq:app_comm_a1} \\
    -i \left\langle \left[ a, \varepsilon(t)e^{i\Delta_a t}a^\dagger - \varepsilon^*(t)e^{-i\Delta_a t}a \right] \right\rangle &= -\varepsilon(t)e^{i\Delta_a t}. \label{eq:app_comm_a2}
\end{align}
Similarly, applying the dual dissipator relation in Eq.~\eqref{eq:app_dissipator_dual} to the decay and thermal excitation channels gives
\begin{align}
    \gamma\bigl(N_b(T) + 1\bigr)\operatorname{Tr}\bigl(a \mathcal{D}_A^{[a]}[\tilde{\rho}_{AB}]\bigr) &= -\frac{\gamma\bigl(N_b(T) + 1\bigr)}{2}\langle a \rangle, \label{eq:app_diss_a1} \\
    \gamma N_b(T)\operatorname{Tr}\bigl(a \mathcal{D}_A^{[a^\dagger]}[\tilde{\rho}_{AB}]\bigr) &= +\frac{\gamma N_b(T)}{2}\langle a \rangle. \label{eq:app_diss_a2}
\end{align}
Summing these yields the equation of motion for the charger expectation value $\langle a \rangle$. Repeating the same procedure for the battery mode $b$ (which is free from direct drive and dissipation) yields the complete set of first-order equations:
\begin{align}
    \frac{d}{dt}\langle a \rangle &= -i g_0\lambda(t)e^{i\Delta' t}\langle b \rangle - \varepsilon(t)e^{i\Delta_a t} - \frac{\gamma}{2}\langle a \rangle, \label{eq:app_dt_a} \\
    \frac{d}{dt}\langle b \rangle &= -i g_0\lambda(t)e^{-i\Delta' t}\langle a \rangle. \label{eq:app_dt_b}
\end{align}
\subsection{Second-Order Moment and Covariance Dynamics}
Because the Hamiltonian is at most bilinear in the creation and annihilation operators and the jump operators are linear, the time derivatives of second-order expectation values couple exclusively to moments of order $\le 2$, forming an exact, closed linear dynamical system.
By systematically applying Eqs.~\eqref{eq:app_comm_identity} and \eqref{eq:app_dissipator_dual} to all second-order operator pairings, we obtain the remaining equations of motion:
\begin{align}
    \frac{d}{dt}\langle a^\dagger a \rangle &= i g_0\lambda(t)\left[ e^{-i\Delta' t}\langle a b^\dagger \rangle - e^{i\Delta' t}\langle a^\dagger b \rangle \right] \nonumber \\
    &\quad - \varepsilon(t)e^{i\Delta_a t}\langle a^\dagger \rangle - \varepsilon^*(t)e^{-i\Delta_a t}\langle a \rangle \nonumber \\
    &\quad - \gamma\langle a^\dagger a \rangle + \gamma N_b(T), \label{eq:app_dt_adaga} \\
    \frac{d}{dt}\langle b^\dagger b \rangle &= i g_0\lambda(t)\left[ e^{i\Delta' t}\langle a^\dagger b \rangle - e^{-i\Delta' t}\langle a b^\dagger \rangle \right], \label{eq:app_dt_bdagb} \\
    \frac{d}{dt}\langle a b^\dagger \rangle &= i g_0\lambda(t)e^{i\Delta' t}\left( \langle a^\dagger a \rangle - \langle b^\dagger b \rangle \right) \nonumber \\
    &\quad - \varepsilon(t)e^{i\Delta_a t}\langle b^\dagger \rangle - \frac{\gamma}{2}\langle a b^\dagger \rangle, \label{eq:app_dt_abdag} \\
    \frac{d}{dt}\langle a^2 \rangle &= -2i g_0\lambda(t)e^{i\Delta' t}\langle a b \rangle - 2\varepsilon(t)e^{i\Delta_a t}\langle a \rangle - \gamma\langle a^2 \rangle, \label{eq:app_dt_a2} \\
    \frac{d}{dt}\langle b^2 \rangle &= -2i g_0\lambda(t)e^{-i\Delta' t}\langle a b \rangle, \label{eq:app_dt_b2} \\
    \frac{d}{dt}\langle a b \rangle &= -i g_0\lambda(t)\left( e^{i\Delta' t}\langle b^2 \rangle + e^{-i\Delta' t}\langle a^2 \rangle \right) \nonumber \\
    &\quad - \varepsilon(t)e^{i\Delta_a t}\langle b \rangle - \frac{\gamma}{2}\langle a b \rangle. \label{eq:app_dt_ab}
\end{align}
Assuming the bipartite system is initially prepared in the global vacuum state $|0, 0\rangle$, the initial conditions are simply given by:
\begin{equation}
    \langle X \rangle(0) = 0, \quad \forall X \in \{a, b, ab, a^\dagger b, a^2, b^2, a^\dagger a, b^\dagger b\}.
\end{equation}
					 
\section{Symplectic Derivation of Passive State Energy and Invariant Covariance}
\label{app:ergotropy_proof}

In this Appendix, we provide a rigorous derivation of the passive energy and the covariance-based invariants using Williamson's theorem for continuous-variable Gaussian states.

\subsection{Gaussian State Representation and Passivity}
Any single-mode Gaussian state $\rho_G$ can be parameterized as a displaced and squeezed thermal state~\cite{Williamson1936,Weedbrook2012}:
\begin{equation}
    \rho_G = D(\alpha) S(\xi) \rho_{\bar{\beta}} S^\dagger(\xi) D^\dagger(\alpha),
    \label{eq:gaussian_decomp}
\end{equation}
where $D(\alpha) = \exp(\alpha b^\dagger - \alpha^* b)$ is the displacement operator, $S(\xi) = \exp[\tfrac{1}{2}(\xi^* b^2 - \xi b^{\dagger 2})]$ is the single-mode squeezing operator with $\xi = r e^{i\theta}$, and
\begin{equation}
    \rho_{\bar{\beta}} = (1 - e^{-\bar{\beta}\omega_0}) \sum_{n=0}^\infty e^{-n \bar{\beta}\omega_0} |n\rangle\langle n|
\end{equation}
is a thermal state at an effective inverse temperature $\bar{\beta}$.

Since unitary transformations preserve the spectrum of $\rho_G$, the passive state within its unitary orbit is uniquely obtained by rearranging the eigenvalues in decreasing order onto the energy eigenbasis. For a single bosonic mode, this passive representative is precisely a thermal state $\rho_{\bar{\beta}}$ with the same symplectic eigenvalue as $\rho_G$~\cite{d40}. Its passive energy is therefore
\begin{equation}
    E_B^{(p)} = \operatorname{Tr}\!\left[ H_B \rho_{\bar{\beta}} \right] = \omega_0 n_B^{(p)}, 
    \qquad 
    n_B^{(p)} = \frac{1}{e^{\bar{\beta}\omega_0} - 1}.
    \label{eq:passive_energy_def}
\end{equation}

\subsection{Williamson's Theorem and the Symplectic Invariant}
We define the canonical, dimensionless quadrature operators for the battery mode $B$ as
\begin{equation}
    x_B = \frac{b + b^\dagger}{\sqrt{2}}, 
    \qquad 
    p_B = \frac{b - b^\dagger}{\sqrt{2}i},
    \label{eq:quadratures}
\end{equation}
satisfying $[x_B, p_B] = i$ (with $\hbar = 1$). Introducing $\Delta \mathbf{r} = (\Delta x_B, \Delta p_B)^T$ with $\Delta r_j \equiv r_j - \langle r_j \rangle$, the covariance matrix $\boldsymbol{\sigma}_B$ is defined by
\begin{equation}
    [\boldsymbol{\sigma}_B]_{jk} = \frac{1}{2}\left\langle \Delta r_j \Delta r_k + \Delta r_k \Delta r_j \right\rangle, 
    \qquad j,k \in \{1,2\}.
    \label{eq:cov_matrix_def}
\end{equation}
By Williamson's theorem~\cite{Williamson1936}, any strictly positive-definite covariance matrix can be diagonalized by a symplectic transformation $\mathbb{S}\in \mathrm{Sp}(2,\mathbb{R})$:
\begin{equation}
    \mathbb{S}\boldsymbol{\sigma}_B\mathbb{S}^T
    = \nu_B \mathbb{I}_{2\times 2}
    = \begin{pmatrix} \nu_B & 0 \\ 0 & \nu_B \end{pmatrix},
    \label{eq:williamson_form}
\end{equation}
where $\nu_B \ge 1/2$ is the symplectic eigenvalue and $\mathbb{S}$ satisfies $\mathbb{S}\boldsymbol{\Omega}\mathbb{S}^T=\boldsymbol{\Omega}$ with $\boldsymbol{\Omega}=\begin{pmatrix}0&1\\-1&0\end{pmatrix}$.
Taking determinants in Eq.~\eqref{eq:williamson_form} and using $\det\mathbb{S}=1$ yields the symplectic invariant
\begin{equation}
    \det \boldsymbol{\sigma}_B = \nu_B^2
    \quad \Rightarrow \quad 
    \nu_B = \sqrt{\det \boldsymbol{\sigma}_B}.
    \label{eq:symplectic_eval}
\end{equation}
For the thermal reference state $\rho_{\bar{\beta}}$, one has $\boldsymbol{\sigma}_{\bar{\beta}}=(n_B^{(p)}+1/2)\mathbb{I}_{2\times 2}$, hence
\begin{equation}
    \nu_B = n_B^{(p)} + \frac{1}{2}
    \quad \Rightarrow \quad 
    n_B^{(p)} = \sqrt{\det \boldsymbol{\sigma}_B} - \frac{1}{2}.
    \label{eq:passive_occ_symp}
\end{equation}

\subsection{Evaluation via Second-Order Ladder Moments}
To connect $\det \boldsymbol{\sigma}_B$ to the dynamical moment equations, we express the covariance entries in terms of ladder-operator moments:
\begin{align}
    \sigma_{xx} &= \frac{1}{2} + \langle b^\dagger b \rangle - |\langle b \rangle|^2 
    + \operatorname{Re}\!\left( \langle b^2 \rangle - \langle b \rangle^2 \right), \\
    \sigma_{pp} &= \frac{1}{2} + \langle b^\dagger b \rangle - |\langle b \rangle|^2 
    - \operatorname{Re}\!\left( \langle b^2 \rangle - \langle b \rangle^2 \right), \\
    \sigma_{xp} &= \operatorname{Im}\!\left( \langle b^2 \rangle - \langle b \rangle^2 \right).
\end{align}
Define the centered quantities
\begin{equation}
    u \equiv \langle b^\dagger b \rangle - |\langle b \rangle|^2, 
    \qquad 
    v \equiv \langle b^2 \rangle - \langle b \rangle^2.
\end{equation}
Then $\sigma_{xx} = u + 1/2 + \operatorname{Re}(v)$, $\sigma_{pp} = u + 1/2 - \operatorname{Re}(v)$, and $\sigma_{xp} = \operatorname{Im}(v)$, giving
\begin{align}
    \det \boldsymbol{\sigma}_B 
    &= \sigma_{xx}\sigma_{pp} - \sigma_{xp}^2 \nonumber \\
    &= \left( u + \frac{1}{2} \right)^2 - |v|^2 \nonumber \\
    &= \frac{1}{4}\left[(1+2u)^2 - 4|v|^2\right] \equiv \frac{1}{4} M(\tau),
    \label{eq:det_to_M}
\end{align}
where $M(\tau)\ge 1$ follows from the Robertson--Schr\"odinger uncertainty relation.
Substituting $\det \boldsymbol{\sigma}_B = M(\tau)/4$ into Eq.~\eqref{eq:passive_occ_symp} yields
\begin{equation}
    n_B^{(p)}(\tau) = \frac{1}{2}\left(\sqrt{M(\tau)} - 1\right),
    \label{eq:passive_occupation_final}
\end{equation}
and therefore the ergotropy
\begin{equation}
    \mathcal{E}_B(\tau) = \omega_0\left(\langle b^\dagger b \rangle(\tau) - n_B^{(p)}(\tau)\right).
\end{equation}

\section{Proof of the Dynamical Decomposition Relation}
\label{app:proof_decomposition}

In this Appendix, we establish an exact superposition property for the expectation value of any system observable $\hat{\mathcal{O}}$ that is at most quadratic in the bosonic field operators. Specifically, we prove
\begin{equation}
    \mathcal{R}_{\hat{\mathcal{O}}}(t; F, T) 
    = \mathcal{R}_{\hat{\mathcal{O}}}^{\mathrm{th}}(t; T) 
    + \mathcal{R}_{\hat{\mathcal{O}}}^{\mathrm{coh}}(t; F),
    \label{eq:app_response_decomp}
\end{equation}
where
\begin{equation}
    \mathcal{R}_{\hat{\mathcal{O}}}(t; F, T) \equiv \operatorname{Tr}\!\left[ \hat{\mathcal{O}}\,\tilde{\rho}_{AB}(t; F, T) \right].
\end{equation}
Here, $\mathcal{R}_{\hat{\mathcal{O}}}^{\mathrm{th}}(t; T)$ denotes the purely thermal response induced by the bath at temperature $T$ in the absence of coherent driving ($F=0$), while $\mathcal{R}_{\hat{\mathcal{O}}}^{\mathrm{coh}}(t; F)$ is the coherent response generated by the external control field $F\equiv\{\varepsilon(t)\}$ at zero temperature ($T=0$).

The proof follows from the affine, block-structured hierarchy of moment equations derived from the Lindblad master equation: because the Hamiltonian is at most bilinear and the jump operators are linear, the equations close exactly at second order, and the only temperature-dependent injection appears as a constant source term proportional to $\gamma N_b(T)$ in the population sector. We collect the first- and second-order moments into
\begin{equation}
    \mathbf{X}_1(t) = \begin{pmatrix} \langle a \rangle \\ \langle b \rangle \end{pmatrix},
    \qquad
    \mathbf{X}_2(t) = \begin{pmatrix} 
        \langle a^\dagger a \rangle \\ 
        \langle b^\dagger b \rangle \\ 
        \langle a b^\dagger \rangle \\ 
        \langle a^\dagger b \rangle \\ 
        \langle a^2 \rangle \\ 
        \langle b^2 \rangle \\ 
        \langle a b \rangle 
    \end{pmatrix}.
\end{equation}
The dynamics takes the affine form
\begin{align}
    \dot{\mathbf{X}}_1(t) &= \mathbf{A}_1(t)\mathbf{X}_1(t) + \mathbf{s}_F(t), \label{eq:X1_affine} \\
    \dot{\mathbf{X}}_2(t) &= \mathbf{A}_2(t)\mathbf{X}_2(t) + \mathbf{K}_F(t)\mathbf{X}_1(t) + \mathbf{s}_T, \label{eq:X2_affine}
\end{align}
where $\mathbf{A}_1(t)$ and $\mathbf{A}_2(t)$ are independent of $\varepsilon(t)$ and $N_b(T)$ (they depend only on $g(t)$ and $\gamma$), $\mathbf{s}_F(t)\propto \varepsilon(t)$ is purely coherent, and $\mathbf{s}_T\propto \gamma N_b(T)\hat{\mathbf{e}}_1$ vanishes at $T=0$. The coupling $\mathbf{K}_F(t)$ is linear in $\varepsilon(t)$ and accounts for the fact that the coherent drive enters second-order moments only through the induced first-order displacements.

With vacuum (zero-displacement) initial conditions,
\begin{equation}
    \mathbf{X}_1(0) = \mathbf{0}, \qquad \mathbf{X}_2(0) = \mathbf{0},
\end{equation}
the unique solution of Eq.~\eqref{eq:X1_affine} is
\begin{equation}
    \mathbf{X}_1(t) = \int_0^t \mathbf{\Phi}_1(t, s)\mathbf{s}_F(s)\,ds,
    \label{eq:X1_solution}
\end{equation}
with $\mathbf{\Phi}_1(t, s)=\mathcal{T}\exp\!\left(\int_s^t \mathbf{A}_1(s')ds'\right)$. Hence,
\begin{equation}
    \left.\mathbf{X}_1(t)\right|_{F=0,\, T} = \mathbf{0},
    \qquad
    \mathbf{X}_1(t) = \left.\mathbf{X}_1(t)\right|_{F,\, T=0}.
\end{equation}
Applying the Duhamel formula to Eq.~\eqref{eq:X2_affine} gives
\begin{equation}
    \mathbf{X}_2(t) = \int_0^t \mathbf{\Phi}_2(t, s)\left[\mathbf{K}_F(s)\mathbf{X}_1(s) + \mathbf{s}_T\right]ds,
\end{equation}
where $\mathbf{\Phi}_2(t, s)=\mathcal{T}\exp\!\left(\int_s^t \mathbf{A}_2(s')ds'\right)$ is independent of $F$ and $T$. By linearity,
\begin{equation}
    \mathbf{X}_2(t)
    =
    \underbrace{\int_0^t \mathbf{\Phi}_2(t, s)\mathbf{s}_T\,ds}_{\text{Thermal contribution}}
    +
    \underbrace{\int_0^t \mathbf{\Phi}_2(t, s)\mathbf{K}_F(s)\mathbf{X}_1(s)\,ds}_{\text{Coherent contribution}}.
    \label{eq:X2_split}
\end{equation}
The two limiting cases follow immediately:
(i) for $F=0$, $\mathbf{s}_F=\mathbf{0}$ and $\mathbf{K}_F=\mathbf{0}$, yielding the thermal term only; 
(ii) for $T=0$, $N_b(T)=0$ and hence $\mathbf{s}_T=\mathbf{0}$, yielding the coherent term only. Therefore
\begin{equation}
    \mathbf{X}_2(t) = \left.\mathbf{X}_2(t)\right|_{F=0,\, T} + \left.\mathbf{X}_2(t)\right|_{F,\, T=0}.
\end{equation}
Finally, since any quadratic observable $\hat{\mathcal{O}}$ is a linear functional of the components of $\mathbf{X}_1(t)$ and $\mathbf{X}_2(t)$, linearity of the trace implies Eq.~\eqref{eq:app_response_decomp}, which completes the proof.

\begin{figure}[t]
    \centering
     \includegraphics[width=\columnwidth]{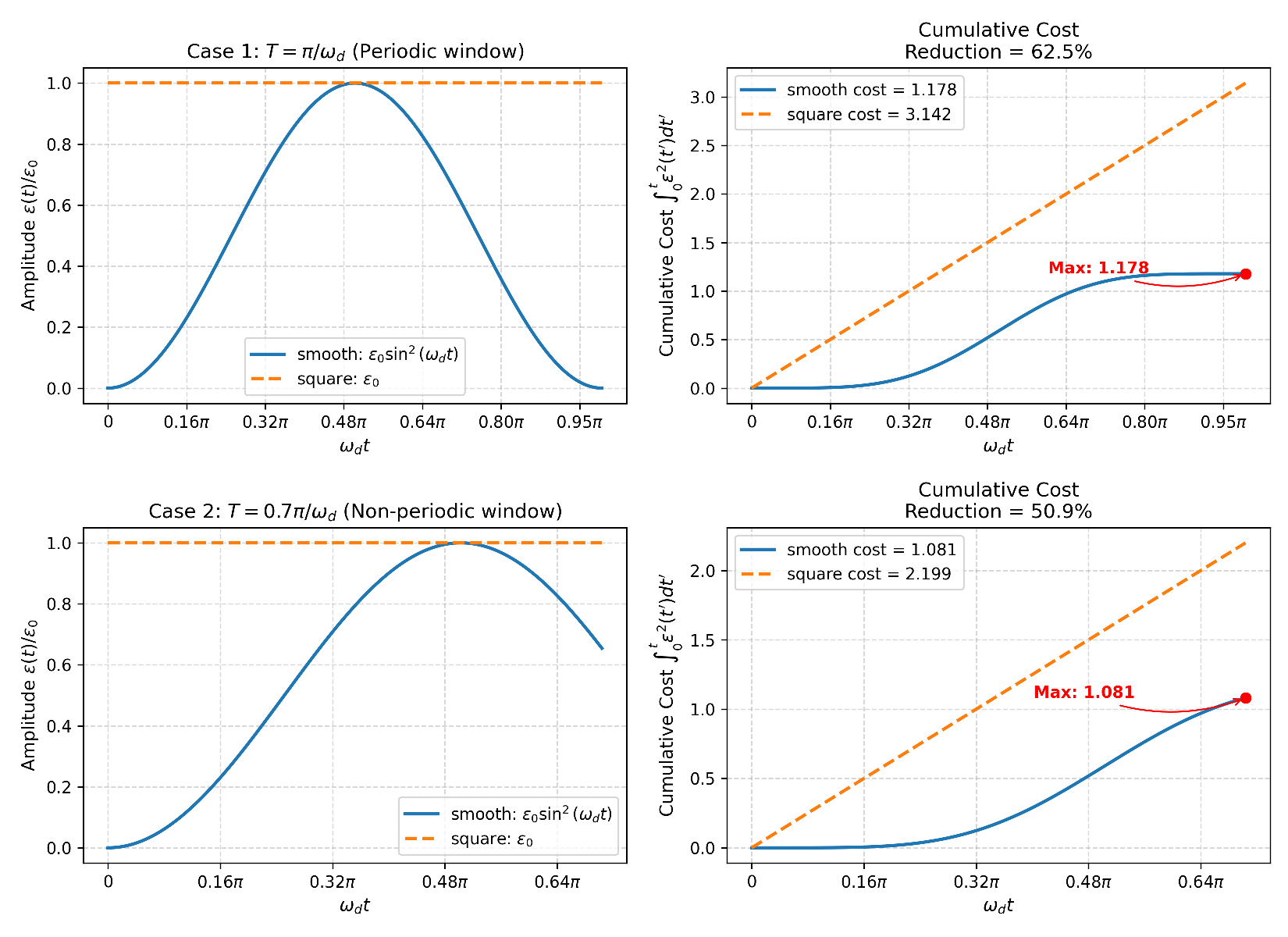}
    \caption{Comparison between smooth envelope driving, $\varepsilon(t)=\varepsilon_0\sin^2(\omega_d t)$, and constant square driving, $\varepsilon(t)=\varepsilon_0$, for two operational time windows. Left panels display the normalized driving amplitude $\varepsilon(t)/\varepsilon_0$ as a function of dimensionless time $\omega_d t$. Right panels trace the cumulative input cost $\int_0^t \varepsilon^2(t')\,dt'$. For the periodic matching window $T=\pi/\omega_d$, the smooth modulation provides an input fluence reduction of $62.5\%$, while for the non-periodic window $T=0.7\pi/\omega_d$, the operational overhead is reduced by $50.9\%$.}
    \label{fig:smooth_square_cost}
\end{figure}
\section{Driving-Cost Analysis}
\label{app:driving-cost}
To benchmark the energetic footprint and control effort across distinct driving profiles, we quantify the fluence via the quadratic driving cost functional over the operational window $\tau$:
\begin{equation}
    C_d \equiv \int_{0}^{\tau} \varepsilon^2(t) \, dt,
    \label{eq:Cd_functional}
\end{equation}
where $\varepsilon(t) \in \mathbb{R}$ represents the time-dependent external driving envelope. The functional $C_d$ serves as a canonical measure of the integrated pulse fluence, parameterizing the resource overhead necessary to synthesize and maintain the control drive field.
\subsection{Analytical Comparison: Square versus Smooth Envelopes}
We contrast a standard rectangular flat-top pulse,
\begin{equation}
    \varepsilon_{\mathrm{sq}}(t) = \varepsilon_0, \qquad t \in [0, \tau],
\end{equation}
against the smoothly modulated envelope
\begin{equation}
    \varepsilon_{\mathrm{sm}}(t) = \varepsilon_0 \sin^2(\omega_d t), \qquad t \in [0, \tau].
    \label{eq:eps_sm}
\end{equation}
The driving cost for the square protocol evaluates directly to
\begin{equation}
    C_{\mathrm{sq}} = \int_0^{\tau} \varepsilon_0^2\,dt = \varepsilon_0^2 \tau.
    \label{eq:Csq}
\end{equation}
For the continuous sine-squared modulation, the cost is given by
\begin{equation}
    C_{\mathrm{sm}} = \varepsilon_0^2 \int_0^{\tau} \sin^4(\omega_d t)\,dt.
    \label{eq:Csm_def}
\end{equation}
Employing the trigonometric power-reduction identity $\sin^4 x = \frac{3}{8} - \frac{1}{2}\cos(2x) + \frac{1}{8}\cos(4x)$, we obtain the closed-form integrated fluence:
\begin{equation}
    C_{\mathrm{sm}} = \varepsilon_0^2 \left[ \frac{3}{8}\tau - \frac{\sin(2\omega_d \tau)}{4\omega_d} + \frac{\sin(4\omega_d \tau)}{32\omega_d} \right].
    \label{eq:Csm_closed}
\end{equation}

We define the relative fluence reduction efficiency as
\begin{equation}
    \eta \equiv 1 - \frac{C_{\mathrm{sm}}}{C_{\mathrm{sq}}}.
    \label{eq:eta_def}
\end{equation}
When the operational window is commensurate with the drive frequency, $\tau = n\pi/\omega_d$ ($n \in \mathbb{Z}^+$), all oscillatory boundary terms identically vanish, yielding the asymptotic ratio
\begin{equation}
    C_{\mathrm{sm}} = \frac{3}{8} C_{\mathrm{sq}} \implies \eta = 1 - \frac{3}{8} = \frac{5}{8} = 62.5\%.
    \label{eq:eta_periodic}
\end{equation}
For non-periodic durations, the boundary terms in Eq.~\eqref{eq:Csm_closed} persist; nevertheless, across the transient charging intervals explored in the main text, the smooth envelope consistently guarantees an energy overhead reduction exceeding $50\%$.
\section{ Derivation of the Instantaneous Steady-State Displacement and the Local Counter-Diabatic Field}
\label{app:appendix_F}
In this Appendix, we present the explicit step-by-step derivation of the instantaneous steady-state displacement amplitude $\bar{\alpha}(t)$ and demonstrate how it determines the local counter-diabatic (CD) driving field. We begin with the Markovian Master Equation for the open system in the rotating frame~\cite{d42, Santos2015, Carmichael1999}:
\begin{equation}
\dot{\rho}(t) = -i[H(t), \rho(t)] + \gamma (N_b + 1) \mathcal{D}[a]\rho(t) + \gamma N_b \mathcal{D}[a^\dagger]\rho(t),
\label{eq:app_master}
\end{equation}
where $\mathcal{D}[o]\rho = o\rho o^\dagger - \frac{1}{2}\{o^\dagger o, \rho\}$, $\gamma$ denotes the spontaneous decay rate, and $N_b$ is the mean thermal photon number of the environment. The effective driving Hamiltonian of the charger mode is given by
\begin{equation}
H(t) = \Delta_a a^\dagger a - i\left(\epsilon(t)a^\dagger - \epsilon^*(t)a\right).
\label{eq:app_H}
\end{equation}
Using the cyclic property of the trace, $\frac{d}{dt}\langle a \rangle = \mathrm{Tr}(a\dot{\rho})$, the equation of motion for the first-order coherent expectation value $\alpha(t) \equiv \langle a(t) \rangle$ is obtained as:
\begin{equation}
\begin{aligned}
\frac{d}{dt}\langle a \rangle &= -i\,\mathrm{Tr}\Big(a [H(t), \rho]\Big) + \gamma (N_b + 1)\,\mathrm{Tr}\Big(a \mathcal{D}[a]\rho\Big) \\
&\quad + \gamma N_b\,\mathrm{Tr}\Big(a \mathcal{D}[a^\dagger]\rho\Big) \\
&= -i\langle [a, H(t)] \rangle - \frac{\gamma}{2}\langle a \rangle.
\end{aligned}
\label{eq:app_exp_a}
\end{equation}
Evaluating the commutator explicitly,
\begin{equation}
[a, H(t)] = \Delta_a [a, a^\dagger a] - i\epsilon(t)[a, a^\dagger] = \Delta_a a - i\epsilon(t).
\end{equation}
Notice that the thermal terms proportional to $N_b$ cancel identically in the first-order moment equation. Substituting back into Eq.~\eqref{eq:app_exp_a} yields the closed differential equation:
\begin{equation}
\dot{\alpha}(t) = \left(i\Delta_a - \frac{\gamma}{2}\right)\alpha(t) + \epsilon(t).
\label{eq:app_diff_alpha}
\end{equation}
Under adiabatic driving, the system instantaneously tracks the steady state of the Liouvillian generator. Setting the rate of change to zero, $\dot{\bar{\alpha}}_{\mathrm{ss}}(t) = 0$, the instantaneous steady-state amplitude $\bar{\alpha}(t)$ satisfies:
\begin{equation}
\left(i\Delta_a - \frac{\gamma}{2}\right)\bar{\alpha}(t) + \epsilon(t) = 0,
\end{equation}
which immediately leads to:
\begin{equation}
\bar{\alpha}(t) = \frac{i\epsilon(t)}{\Delta_a - i\gamma/2}.
\label{eq:app_alpha_steady}
\end{equation}
Using the time-derivative $\dot{\bar{\alpha}}(t)$, the unitary transformation $D[\bar{\alpha}(t)]$ generates the gauge potential that counteracts non-adiabatic excitations. This procedure directly leads to the counter-diabatic Hamiltonian $H_{\mathrm{CD}}(t)$ and the renormalized driving field $\epsilon_{\mathrm{CD}}(t)$, which are presented and analyzed in Sec.~\ref{sec:CD_dynamics} [see Eqs.~\eqref{eq:alpha_steady_state}--\eqref{eq:H_CD_explicit} of the main text].
\section{Exact Biorthogonal Spectrum of the Thermal Liouvillian}
\label{app:biorthogonal_spectra}
In this Appendix, we derive the exact eigenvalue spectrum and right eigenoperators of the thermal Liouvillian superoperator:
\begin{equation}
    \mathcal{L}\rho = -i[\Delta_a a^\dagger a, \rho] + \gamma(N_b + 1)\mathcal{D}[a]\rho + \gamma N_b \mathcal{D}[a^\dagger]\rho,
    \label{eq:app_L_thermal}
\end{equation}
where $\mathcal{D}[L]\rho \equiv L\rho L^\dagger - \frac{1}{2}\{L^\dagger L, \rho\}$ is the standard Lindblad dissipator.

Acting with $\mathcal{L}$ on the outer-product basis $|m\rangle\langle n|$ of the Fock space yields
\begin{equation}
    \begin{aligned}
        \mathcal{L}|m\rangle\langle n| &= \left[ -i\Delta_a(m - n) - \frac{\gamma}{2}(2N_b + 1)(m + n) - \gamma N_b \right] |m\rangle\langle n| \\
        &\quad + \gamma(N_b + 1)\sqrt{mn}\,|m - 1\rangle\langle n - 1| \\
        &\quad + \gamma N_b\sqrt{(m + 1)(n + 1)}\,|m + 1\rangle\langle n + 1|.
    \end{aligned}
\end{equation}
For a fixed coherence index $j \equiv m - n \in \mathbb{Z}$, the linear subspace
\begin{equation}
    \mathcal{H}_j = \operatorname{span}\left\{ |k\rangle\langle k + j| \right\}_{k=0}^\infty
\end{equation}
forms an invariant manifold under the action of $\mathcal{L}$. We expand the right eigenoperators in the form
\begin{equation}
    \rho_j = \sum_{k=0}^{\infty} c_k^{(j)} |k\rangle\langle k + j|, \qquad \mathcal{L}\rho_j = \Lambda\,\rho_j.
    \label{eq:app_rhoj_ansatz}
\end{equation}
Substitution into the eigenvalue equation leads to the second-order difference relation:
\begin{equation}
    \begin{aligned}
        &\gamma(N_b + 1)\sqrt{k(k + j)}\,c_{k-1}^{(j)} \\
        &+ \left[ -i\Delta_a j - \frac{\gamma}{2}(2N_b + 1)(2k + j) - \gamma N_b - \Lambda \right] c_k^{(j)} \\
        &+ \gamma N_b \sqrt{(k + 1)(k + j + 1)}\,c_{k+1}^{(j)} = 0.
    \end{aligned}
    \label{eq:app_ck_recurrence}
\end{equation}
To eliminate the combinatorial square-root weights, we apply the substitution
\begin{equation}
    c_k^{(j)} = \sqrt{\frac{k!}{(k + j)!}}\,Q_k^{(j)},
    \label{eq:app_Q_def}
\end{equation}
transforming Eq.~\eqref{eq:app_ck_recurrence} into
\begin{align}
    &\gamma(N_b + 1) k\, Q_{k-1}^{(j)} \nonumber \\
    &\quad + \left[ -i\Delta_a j - \frac{\gamma}{2}(2N_b + 1)(2k + j) - \gamma N_b - \Lambda \right] Q_k^{(j)} \nonumber \\
    &\quad + \gamma N_b (k + j + 1) Q_{k+1}^{(j)} = 0.
    \label{eq:F7}
\end{align}
We define the corresponding generating function
\begin{equation}
    \Phi_j(z) = \sum_{k=0}^{\infty} Q_k^{(j)} z^k.
    \label{eq:app_generating_function}
\end{equation}
Multiplying Eq.~\eqref{eq:F7} by $z^k$ and summing over $k \ge 0$ yields the first-order ordinary differential equation:
\begin{align}
    &\left[ \gamma N_b z^2 - \gamma(2N_b + 1)z + \gamma(N_b + 1) \right] \frac{d\Phi_j(z)}{dz} \nonumber \\
    &\quad + \left[ \gamma N_b (j + 1)z - \frac{\gamma}{2}(2N_b + 1)j - \gamma N_b - i\Delta_a j - \Lambda \right] \Phi_j(z) = 0.
    \label{eq:F9}
\end{align}
The characteristic polynomial in front of the derivative factorizes into real roots:
\begin{equation}
    \gamma N_b (z - z_1)(z - z_2), \qquad z_1 = 1, \quad z_2 = 1 + \frac{1}{N_b}.
    \label{eq:roots}
\end{equation}
Decomposing the logarithmic derivative into partial fractions,
\begin{equation}
    \frac{\Phi_j'(z)}{\Phi_j(z)} = \frac{\mu_1}{z - z_1} + \frac{\mu_2}{z - z_2},
\end{equation}
yields the respective residues:
\begin{align}
    \mu_1 &= \frac{\Lambda - i\Delta_a j + \frac{\gamma}{2}j}{\gamma}, \label{eq:mu1} \\
    \mu_2 &= -(j + 1) - \mu_1 \nonumber \\
          &= -\frac{\Lambda - i\Delta_a j + \frac{\gamma}{2}j}{\gamma} - (j + 1). \label{eq:mu2}
\end{align}
Integrating directly gives the general solution:
\begin{equation}
    \Phi_j(z) = \mathcal{C} \left( 1 - \frac{z}{z_1} \right)^{\mu_1} \left( 1 - \frac{z}{z_2} \right)^{\mu_2}.
    \label{eq:Phi_solution}
\end{equation}
Physical trace-class density operators require the generating function $\Phi_j(z)$ to remain regular within the unit disk $|z| \le 1$. The boundary singular point at $z_1 = 1$ is removable if and only if the exponent terminates, reducing the non-analytic factor to an exact polynomial degree. Setting
\begin{equation}
    \mu_1 = k, \qquad k = 0, 1, 2, \dots
\end{equation}
directly establishes the quantization condition for the complete Liouvillian spectrum:
\begin{equation}
    \Lambda_{j, k} = i\Delta_a j - \gamma\left( \frac{|j|}{2} + k \right), \qquad j \in \mathbb{Z}, \quad k \in \{0, 1, 2, \dots\}.
    \label{eq:exact_spectrum}
\end{equation}
The polynomial termination in $\Phi_j(z)$ reveals that the radial expansion coefficients $Q_k^{(j)}$ map directly onto the generalized Laguerre polynomials $L_k^{(j)}$.

The unique null eigenvalue $\Lambda_{0, 0} = 0$ corresponds to the thermal equilibrium steady state ($j=0, k=0$):
\begin{equation}
    \rho_{\mathrm{ss}} = \frac{1}{N_b + 1} \sum_{n=0}^{\infty} \left( \frac{N_b}{N_b + 1} \right)^n |n\rangle\langle n|.
    \label{eq:steady_state}
\end{equation}
Equation~\eqref{eq:exact_spectrum} rigorously proves that the decay rates $\operatorname{Re}(\Lambda_{j, k})$ and transition oscillation frequencies $\operatorname{Im}(\Lambda_{j, k})$ are fundamentally invariant with respect to the thermal population $N_b$, confirming that thermal noise alters only the state representations and not the underlying Liouvillian spectrum.

\end{document}